\documentclass[fleqn,usenatbib]{mnras}

\usepackage{color}
\usepackage[T1]{fontenc}
\usepackage{ae,aecompl}
\usepackage{xcolor}
\usepackage{ulem}

\usepackage{graphicx}	
\usepackage{amsmath}	
\usepackage{cases}
\usepackage{adjustbox}

\usepackage{graphicx}	
\usepackage{amssymb}	
\usepackage{newtxtext,newtxmath}
\usepackage{cases}
\usepackage{array,multirow}
\usepackage{upgreek}
\usepackage{textcomp}
\usepackage{rotating}

\newcommand{\Msol}{\;{\rm M}_{\odot}}
\newcommand{\Rsol}{\;{\rm R}_{\odot}}
\newcommand{\gram}{\;\mathrm{g}}
\newcommand{\cm}{\;\mathrm{cm}}

\newcommand{\pc}{\;\mathrm{pc}}

\newcommand{\rstar}{\;R_{\star}}
\newcommand{\mstar}{\;M_{\star}}
\newcommand{\Mbh}{M_{\bullet}}

\newcommand{\mesa}{{\small MESA}}

\usepackage{tikz}

\title[Red giant collision in nuclei]{Collisions of red giants in galactic nuclei}

\author[T. Ryu et al.]{
Taeho Ryu$^{1,2}$,\thanks{E-mail: tryu@mpa-garching.mpg.de}
Pau Amaro Seoane$^{3,\,4,\,5,\,6}$,
Andrew~M.~Taylor$^{7}$,
Sebastian~T. Ohlmann$^{8}$,
\vspace*{0.1cm}\\%
$^{1}$ Max Planck Institute for Astrophysics, Karl-Schwarzschild-Str.~1, 85748 Garching, Germany\\%
$^{2}$ Physics and Astronomy Department, Johns Hopkins University, Baltimore, MD 21218, USA\\%
$^{3}$Universitat Politècnica de València, Spain\\
$^{4}$Max Planck Institute for Extraterrestrial Physics, Garching, Germany\\
$^{5}$Higgs Centre for Theoretical Physics, Edinburgh, UK\\
$^{6}$Kavli Institute for Astronomy and Astrophysics, Beijing 100871, China\\
$^{7}$DESY, Zeuthen, Germany\\
$^{8}$Max Planck Computing and Data Facility, Garching, Germany
}

\date{Accepted XXX. Received YYY; in original form ZZZ}

\pubyear{2022}

\begin{document}
\label{firstpage}
\pagerange{\pageref{firstpage}--\pageref{lastpage}}
\maketitle

\begin{abstract}
In stellar-dense environments, stars can collide with each other. For collisions close to a supermassive black hole (SMBH), the collisional kinetic energy can be so large that the colliding stars can be destroyed, potentially releasing an amount of energy comparable to that of a supernova. These black hole-driven disruptive collisions (BDCs) have been examined mostly analytically, with the non-linear hydrodynamical effects being left largely unstudied. Using the moving-mesh hydrodynamics code {\small AREPO}, we investigate high-velocity ($>10^{3}$ km/s) collisions between 1M$_{\odot}$ giants with varying radii, impact parameters, and initial approaching velocities, and estimate their observables. Very strong shocks across the collision surface efficiently convert $\gtrsim10\%$ of the initial kinetic energy into radiation energy. The outcome is a gas cloud expanding supersonically, homologously, and quasi-spherically, generating a flare with a peak luminosity $\simeq 10^{41}-10^{44}$ erg/s in the extreme UV band ($\simeq 10$ eV). The luminosity decreases approximately following a power-law $t^{-0.7}$ initially, then $t^{-0.4}$ after $t\simeq$10 days at which point it would be bright in the optical band ($\lesssim 1$eV). Subsequent, and possibly even brighter, emission would be generated due to the accretion of the gas cloud onto the nearby SMBH, possibly lasting up to multi-year timescales. This inevitable BH-collision product interaction can contribute to the growth of BHs at all mass scales, in particular, seed BHs at high redshifts. Furthermore, the proximity of the events to the central BH makes them a potential tool for probing the existence of dormant BHs, even very massive ones which cannot be probed by tidal disruption events.
\end{abstract}

\begin{keywords}

\end{keywords}



\section{Introduction}\label{sec:intro}

Dynamical interactions between stars in stellar-dense environments, e.g., globular clusters and galactic centers, play a crucial role in driving the evolution of the host and determining its thermodynamic state \citep{Hut1992}.  If the stellar density is sufficiently high, stars can collide with relative velocities comparable to the dispersion velocity of the host. In globular clusters, up to 40\% of main-sequence stars in the core would undergo a collision during the lifetime of the cluster \citep{HillsDay1976}. For clusters with very high number densities ($\gtrsim 10^{7}\pc^{-3}$), a star may suffer multiple such collisions \citep{DaleDavies2006}. 

Galactic centers are extreme environments where stars are densely packed (e.g., $10^{6}-10^{7}\pc^{-3}$ for nuclear clusters,  \citealt{Neumayer+2020} and references therein) around a supermassive black hole (SMBH). Because the relative velocity between stars near the SMBH is roughly the Keplerian speed\footnote{\citet{Sellgren+1990} observed a decrease in the 
CO absorption line strength in the central region of our Galaxy, confirming that the velocity dispersion of stars increases toward the center.} $\propto r^{-0.5}$, stars near the BH would collide at very high speeds (e.g., $v_{\rm rel}\gtrsim 2000$km/s within $\simeq 0.1\pc$ around a $10^{7}\Msol$ BH). If the kinetic energy of the collision ($\gtrsim10^{50}$~erg for a collision between two stars with mass $\mstar=1\Msol$ and $v_{\rm rel}\gtrsim 2000$ km/s) is greater than the binding energy of the stars ($10^{48}-10^{49}$~erg for $\mstar=1\Msol$), the stars would be destroyed, leaving behind an expanding gas cloud. If even a small fraction of the collisional kinetic energy is converted into radiation, the high-velocity collision can generate a bright electromagnetic transient from the Galactic nucleus region. 

The total rates of such events between main-sequence stars have been estimated to be $10^{-4}-10^{-5}$ yr$^{-1}$ galaxy$^{-1}$ \citep{Rose+2020,AmaroSeoane2023, Rose+2023} if the core is fully relaxed to the Bahcall-Wolf density power-law $\propto r^{-7/4}$ \citep{BahcallWolf1976}\footnote{While the Bahcall-Wolf solution is a mathematically correct solution when all stars have the same mass, in the realitistic situation where the stellar mass distribution is inhomogeneous, the slope can be steeper \citep{AlexanderHopman2009,PretoAmaroSeoane2010}.}. The rate for collisions between giants could be higher due to larger cross-sections \citep{AmaroSeoane2023}. 
However, if collisions continuously deplete the inner part of the stellar-density cusp, the rate would become smaller, e.g., $\simeq 10^{-5}-10^{-7}$ yr$^{-1}$ galaxy$^{-1}$ for main-sequence stars, depending on the assumption of the stellar influx into the center \citep{BalbergYassur2023}. Since these powerful collisions essentially destroy stars in galactic center environments, these events can affect the frequency of other types of nuclear transients. For example, \cite{BalbergYassur2023} suggests that high-velocity collisions can almost completely suppress extreme mass-ratio inspirals.

High-velocity collisions between main-sequence stars \citep[e.g.,][]{BenzHills1987,BenzHills21992,Lai+1993,Rauch+1999,FreitagBenz2005} have been studied using numerical simulations, focusing on the mass ejection and the impact of such collisions on the thermodynamic state of the host, rather than their observation signatures. Observational signatures of the electromagnetic radiation from such collisions have been studied mostly analytically. For example,    \citet{Balberg+2013} showed that two stars in a compact binary can collide at high speed when passing very close to a SMBH, which can generate a flare as bright as supernovae. Recently, \citet{AmaroSeoane2023} analytically investigated the observables of high velocity collisions between stars of various types in galactic nuclei. They found that the peak luminosity of high-velocity collisions can be as high as $10^{44}\eta_{\rm rad}$~erg/s. Here, $\eta_{\rm rad}$ is one of the determining factors which measures how efficiently the initial kinetic energy is converted into radiation energy. If $\eta_{\rm rad}$ is of order unity, the peak luminosity can be comparable to different types of nuclear transients, such as tidal disruption events. However, $\eta_{\rm rad}$ in their work was left as a free parameter because evaluating $\eta_{\rm rad}$ involves non-linear hydrodynamics effects such as shocks, which cannot be done analytically.  

In this paper, we investigate the hydrodynamics of high-velocity collisions between $1\Msol$ giants and numerically estimate the radiation conversion efficiency and their observables, using the moving-mesh hydrodynamics code {\small AREPO} \citep{Arepo,Arepo2,ArepoHydro}. In the simulations, we consider collisions with $v_{\rm rel}=10^{4}$~km/s between two identical $1\Msol$ giants with four different radii ($R_{\star} =10\Rsol$, $20\Rsol$, $50\Rsol$, and $100\Rsol$), four impact parameters ($b=0.04\rstar$, $0.2\rstar$, $0.4\rstar$, and $0.8\rstar$), and three initial approaching velocity ($v_{\rm rel}=10^{4}$~km/s, $5\times10^{3}$ km/s, and $2.5\times10^{3}$ km/s). The largest approaching speed corresponds to roughly the largest relative velocity for stellar collisions near the BH, i.e., the Keplerian velocity at the smallest possible distance from the BH where at least two stars exist for a typical stellar density around a massive BH assuming the Bahcall-Wolf power law:  $r\simeq 10^{-5}\pc$ for $10^{5}\Msol$ BH, $\simeq 10^{-4}\pc$ for $10^{6}\Msol$ BH, and $\simeq 10^{-3}\pc$ for $10^{7}\Msol$ BH. Because collisions with lower relative velocities are expected to create fainter transients, our simulations with the largest $v_{\rm rel}$ would provide an upper limit for the luminosity and total radiated energy of these events. 

This paper is organized as follows. We describe our methods in \S~\ref{sec:method}, including the code description (\S~\ref{subsec:code}), stellar models (\S~\ref{subsec:stellarmodel}), and initial conditions (\S~\ref{subsec:initialcondition}). Then we present our results in \S~\ref{sec:result} and discuss astrophysical implications for the collisions in \S~\ref{sec:discussion}. Finally, we summarize and conclude in \S~\ref{sec:summary}.

\section{Methods}\label{sec:method}

\subsection{Code}\label{subsec:code}

We perform a suite of 3D hydrodynamic simulations of high-velocity collisions between red giants using the massively parallel gravity and magnetohydrodynamics moving-mesh code {\small AREPO} \citep{Arepo,ArepoHydro,Arepo2}. The code inherits advantages of the two widely used hydrodynamical schemes, the Lagrangian smoothed particle method and the Eulerian finite-volume method, allowing for an accurate treatment of supersonic flows and shock capturing without introducing an artificial viscosity, and low advection errors. We use the ideal equation of states that takes into account radiation pressure assuming local thermodynamic equilibrium, 
\begin{align}
     P = \frac{\rho k_{\rm B}T}{\mu m_{\rm p}} + \frac{4\sigma}{3c}T^{4},
\end{align}
where $P$ is the total pressure, $\rho$ the density, $k_{\rm B}$ the Boltzmann constant, $T$ the temperature, $\mu=0.62$ the mean molecular weight, $m_{\rm p}$ the proton mass, and $\sigma$ the Stefan-Boltzmann constant.

\begin{figure}
	\centering
	\includegraphics[width=8.6cm]{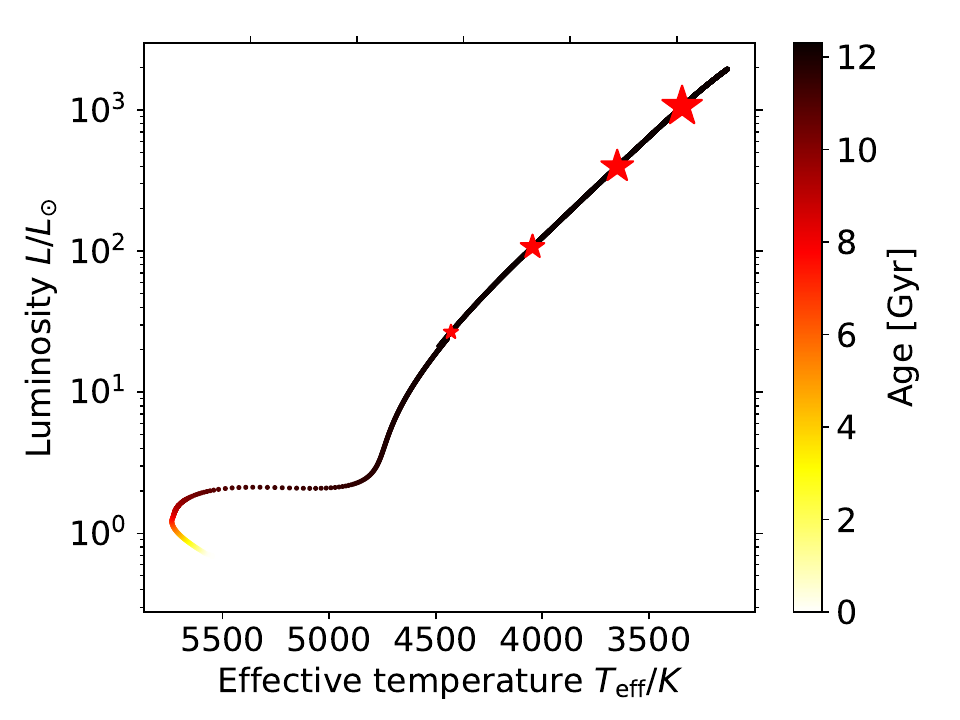}
\caption{Evolution of a $1\Msol$ star in a Hertzsprung–Russell diagram. The color bar shows the age of the star. The four star symbols mark the four giant models adopted for collision experiments: (from smallest to largest symbols) $\rstar\simeq9$, 20, 50, and 100$\Rsol$.}
	\label{fig:HRdiagram}
\end{figure}

\subsection{Stellar model}
\label{subsec:stellarmodel}

We adopt the internal structure of giants evolved using the 1D stellar evolution code \mesa{} \citep[version r22.05.1][]{Paxton+2011,paxton:13} to model giants in 3D. The star has an initial mass $\mstar=1\Msol$ and a metallicity of $Z=0.02$. We treat the mixing processes and winds following \citet{Choi+2016}. More specifically, we model convection using the mixing length theory with a mixing length parameter of $1.81$. We adopt the \citet{ledoux_stellar_1947} criterion to determine the boundary of the convective regions and the exponential overshoot prescription \citep{herwig_evolution_2000} with parameters $f = 0.016$ and $f_0 = 0.008$ at the top of the core and $f = 0.0174$, $f_0 = 0.0087$ at the bottom of the hydrogen-burning shell. Semiconvection is treated following \citet{langer_semiconvective_1983} with an efficiency factor of 0.1. We allow the star on the red giant branch to lose mass via wind following the prescription from \citet{Reimers+1975} with scaling factor of $0.1$.

Figure~\ref{fig:HRdiagram} shows the evolution of the $1\Msol$ star in a Hertzsprung–Russell diagram until it reaches the tip of the red-giant branch. We take the giants at four different evolutionary stages where their radii are $\rstar \simeq 10$, $20$, $50$, and $100 \Rsol$ (indicated by the star symbols in the figure). 

\begin{figure}
	\centering
	\includegraphics[width=8.6cm]{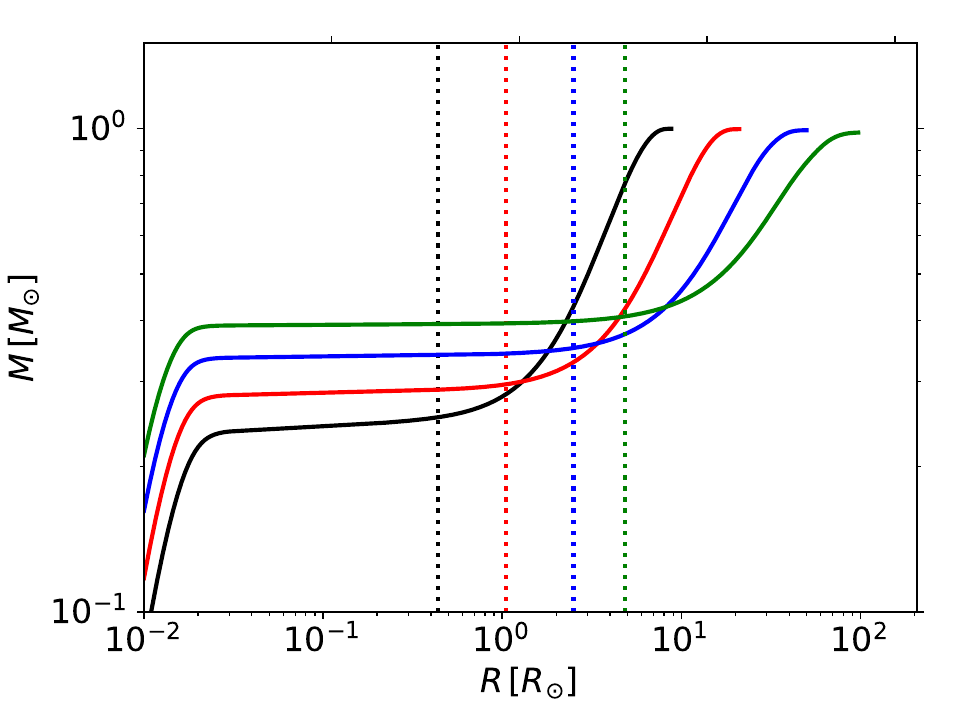}
\caption{Enclosed mass as a function of radius for the four giants with $\rstar=10\Rsol$ (black), $20\Rsol$ (red), $50\Rsol$ (blue), $100\Rsol$ (green). The vertical dotted lines, sharing the same color, indicate the size of the region modelled using a point particle. Although the point particle size is greater than the size of the core ($R\simeq 0.02\Rsol$), given the flat mass-radius relation between the core radius and the point particle radius, we essentially retain the total energy budget inside the star above the core with significantly low computational costs.}
	\label{fig:radius_mass}
\end{figure}

We construct 3D giants from the 1D giant models using the method developed in \citet{Ohlmann+2017} with $10^{6}$ cells. Modeling the entire giant with gas cells is computationally expensive given very steep density gradients. So instead, we model the inner part of the star with a point particle, representing effectively the core. Furthermore, we place gas cells on top of it such that the internal structure above the core matches with the \mesa{} model while the entire star stays in hydrostatic equilibrium. The point particle interacts only gravitationally with gas: it only gravitationally pulls the envelope which is cancelled by the pressure gradient of the gas when the star is in isolation. We choose that the size of the region modelled using a point particle is 5\% of the stellar radius (``point particle radius''). The point particle radius is in fact greater than the size of the core ($R\simeq 0.02\Rsol$). This choice is justified by the fact that the mass of the core is effectively the same as the enclosed mass within $\simeq0.05\rstar$ (vertical dotted lines), as illustrated in Figure~\ref{fig:radius_mass}. This means the total binding energy inside our 3D giants is essentially the same as what we would have had when the point particle radius were exactly the core radius. With this choice of the point radius, while we reduce computational costs significantly, we lose only a  small fraction of the total energy budget inside the star. 

\begin{figure}
	\centering
	\includegraphics[width=8.6cm]{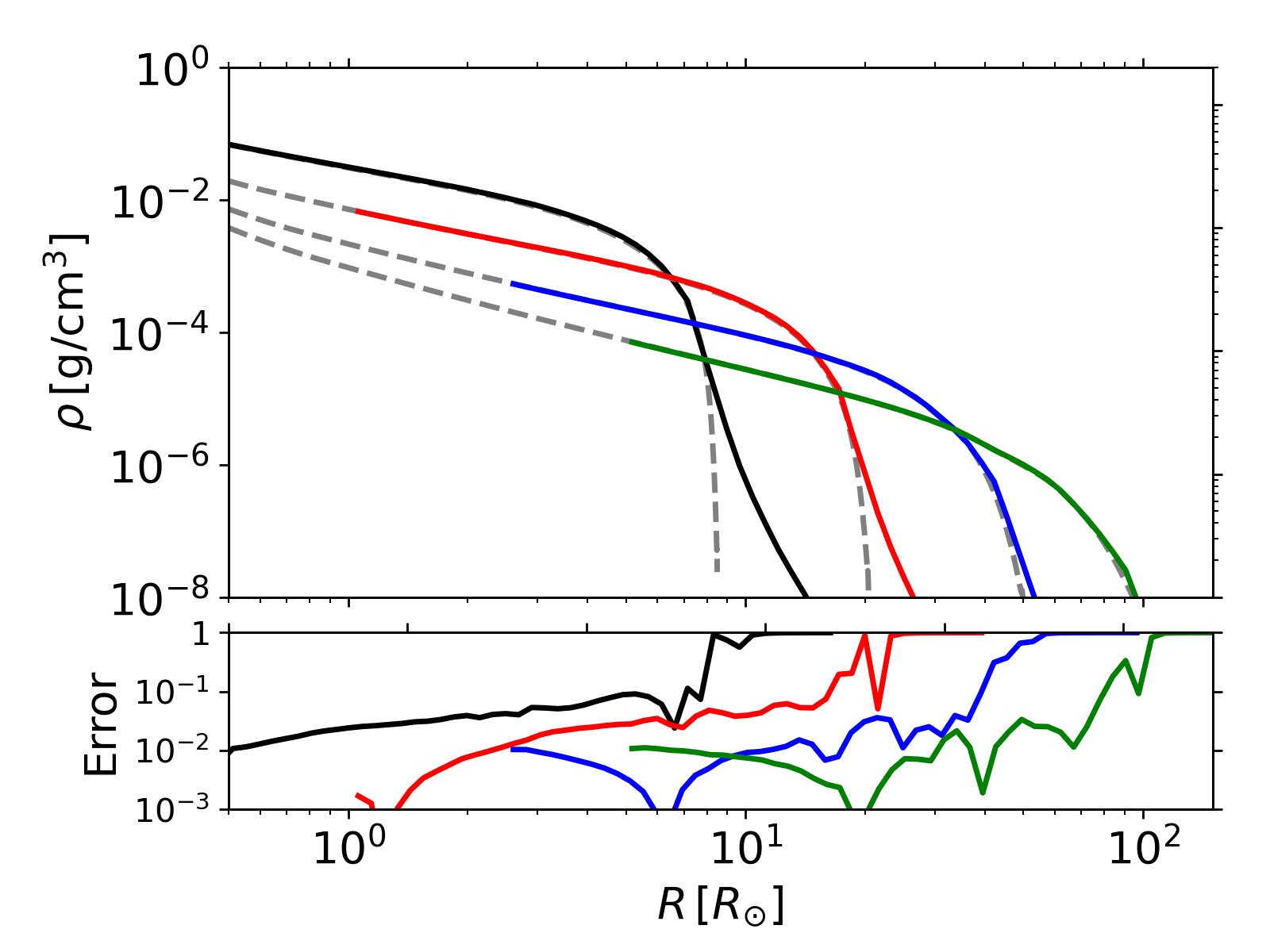}
\caption{The radial density profile (\textit{top}) of the giants with four different radii relaxed for five - ten stellar dynamical times and the relative error with respect to the \mesa{} models (\textit{bottom}), as a function of radius from the core. The dashed grey lines in the \textit{top} panel show the density profiles of the \mesa{} models. The density profiles of the 3D stars match well with the MESA models within a few $\%$ except for those at the stellar surface. }
	\label{fig:density}
\end{figure}

We then relax the 3D stars fully in isolation, which usually takes 5 - 10 stellar dynamical times ($\sqrt{\rstar^{3}/G\mstar}$). 
Figure~\ref{fig:density} shows the radial density of the fully relaxed stars above the point particle (\textit{top} panel) and their errors (\textit{bottom} panel) relative to the \mesa{} models. The relative errors of the density of the inner part of the stars, where most of the mass is concentrated, are less than a few \%. Although the errors at the surface are relatively large, the deviation of such small masses at the surface, corresponding to the plateau at the end of each line in  Figure~\ref{fig:radius_mass}, should not affect our results.

We performed resolution tests for nearly head-on collisions between giants with $\rstar=100\Rsol$ with different resolutions. The choice of the collision parameters are motivated by the fact that the impact of the shock in such a collision is the strongest (see Figure~\ref{fig:eta}), which requires the highest resolution. We first constructed giants with $N=2.5\times10^{5}$, $5\times10^{5}$, $10^{6}$, $2\times10^{6}$ and $4\times10^{6}$ cells and performed the collision experiments. We find that the results have already converged very well when $N\geq10^{6}$: the conversion factor $\eta_{\rm rad}$, defined in Equation~\ref{eq:eta}, differs by less than 1\%. In fact, the difference in $\eta_{\rm rad}$ between $N\leq5\times10^{5}$ and $N=10^{6}$ is already reasonably small, $\lesssim20\%$ for $N=2.5\times 10^{5}$ and $\lesssim 10\%$ for $N=5\times 10^{5}$ relative to the case with $N\geq 10^{6}$. Furthermore, we confirmed that the total energy is conserved within $\lesssim 1\%$ until the end of the simulations.

\begin{table}
\begin{tabular}{ c c c c c} 
\hline
Model number & Mass & Radius  & $v_{\rm rel}$  & Impact parameter $b$ \\
   -  &  $\Msol$  &  $\Rsol$ & 1000~km/s & $\rstar$ \\
\hline
1 & 1 & 10 & 10 & 0.04\\
2 & 1 & 10 & 10 & 0.2\\
3 & 1 & 10 & 10 & 0.4\\
4 & 1 & 10 & 10 & 0.8\\
5 & 1 & 10 & 5 & 0.04\\
6 & 1 & 10 & 2.5 & 0.04\\
7 & 1 & 20 & 10 & 0.04\\
8 & 1 & 50 & 10 & 0.04\\
9 & 1 & 100 & 10 & 0.04\\
\hline
\end{tabular}
\caption{Initial parameters: (from left to right) model number, stellar mass, stellar radius, relative velocity $v_{\rm rel}$ at infinity, and impact parameter $b$. }\label{tab:initialparameter}
\end{table}

\subsection{Initial conditions}\label{subsec:initialcondition}

We place two identical stars, initially separated by $10\rstar$, on a hyperbolic orbit with some relative velocity at infinity $v_{\rm rel}$. So it takes $10\rstar/v_{\rm rel}\simeq (0.1 - 1)$ days, depending on $\rstar$ and $v_{\rm rel}$, until the two stars collide. We note that the time is measured since collision in this paper: accordingly, the initial time of the simulations is $t\simeq -(0.1-1)$ days. Those stars are embedded in a low-density background medium with density of $10^{-18}\gram/\cm^{3}$ and temperature of $10^{4}$ K. The background density is comparable to the density of the interstellar medium at the Galactic center ranging between $10^5$ to
$10^6$ particles per$\cm^{3}$ \citep{Gillessen+2019} at Galactic center distances that dominate the collision rate \citep[see][]{AmaroSeoane2023}. We discuss the impact of the background density and temperature on the properties of collision products in \S\ref{subsec:ism}. Our fiducial model is the near-head-on collision between the two $10\Rsol$ giants initially approaching towards each other at $v_{\rm rel}=10^{4}$km/s with an impact parameter $b=0.04\rstar$. Here, $b=0.04\rstar$ is the smallest possible impact parameter given the softening length of the point particle: in other words,  the gravity of the point particles becomes inaccurate at the closest approach distance with $b<0.04\rstar$. For this giant, we additionally consider off-axis collisions with larger impact parameters, $b=$0.2, 0.4, and 0.8$\rstar$, and two additional $v_{\rm rel} = 2500$ and $5000$ km/s, to study the dependence of the impact parameter and the collision velocity, respectively. For larger giants, we only consider the near head-on collisions with $v_{\rm rel}=10^{4}$ km/s. The initial parameters of the models are summarized in Table~\ref{tab:initialparameter}.

\begin{figure*}
	\centering
\includegraphics[width=17.5cm]{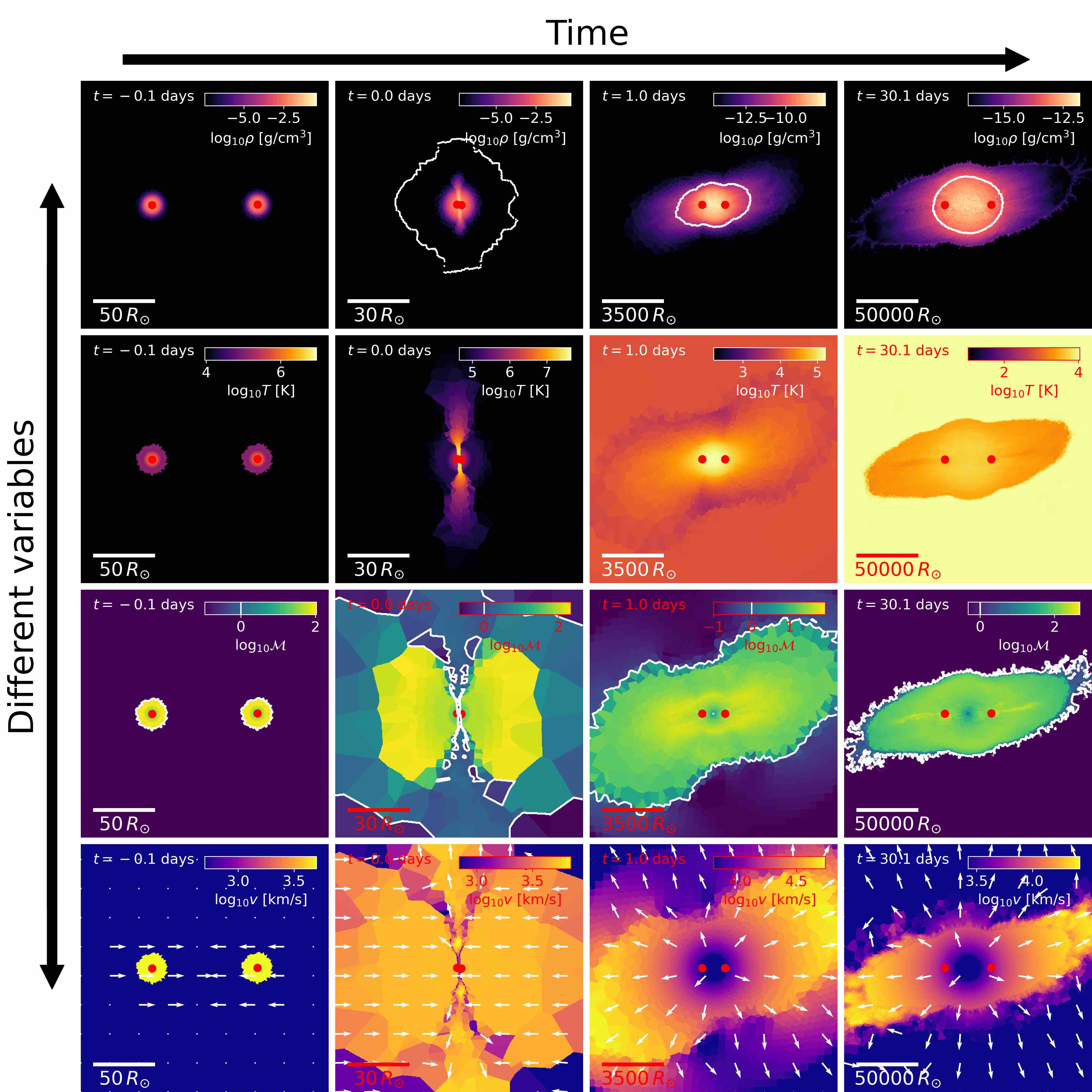}
\caption{Density $\rho$ (\textit{top}), temperature $T$ (\textit{top-middle}), mach number $\mathcal{M}$ (\textit{bottom-middle}), and speed $v$ (\textit{bottom}) of gas in a nearly head-on ($b=0.04R_{\star}$) collision between two giants with $\rstar=10\Rsol$ at four different times, $t=-0.07$ days (before collision), 0 days (at collision), 1 days and 30 days (after collision). The red dots in each panel indicate the location of the cores. The white contour line in the \textit{top} panel for $\rho$ show the location of the photosphere at which the radially integrated optical depth $\simeq 1$ and those in the \textit{bottom-middle} panel for $\mathcal{M}$ the boundaries at $\mathcal{M}\simeq 1$. The arrows in the \textit{bottom} panels indicate the direction of gas motion.  Initially the two stars start to move towards each other with $v_{\rm rel}=10^{4}$ km/s (\textit{left}). At collision, very steep pressure gradients are built up at the collision surface and strong shocks are created when the incoming gas collides with the pressure barrier (\textit{left-middle}). The gas bounces off and expands quasi-spherically and homologously at supersonic speeds (\textit{right-middle } and \textit{right}).   }
	\label{fig:evolution}
\end{figure*}

\begin{figure*}
	\centering
    \includegraphics[width=8.6cm]{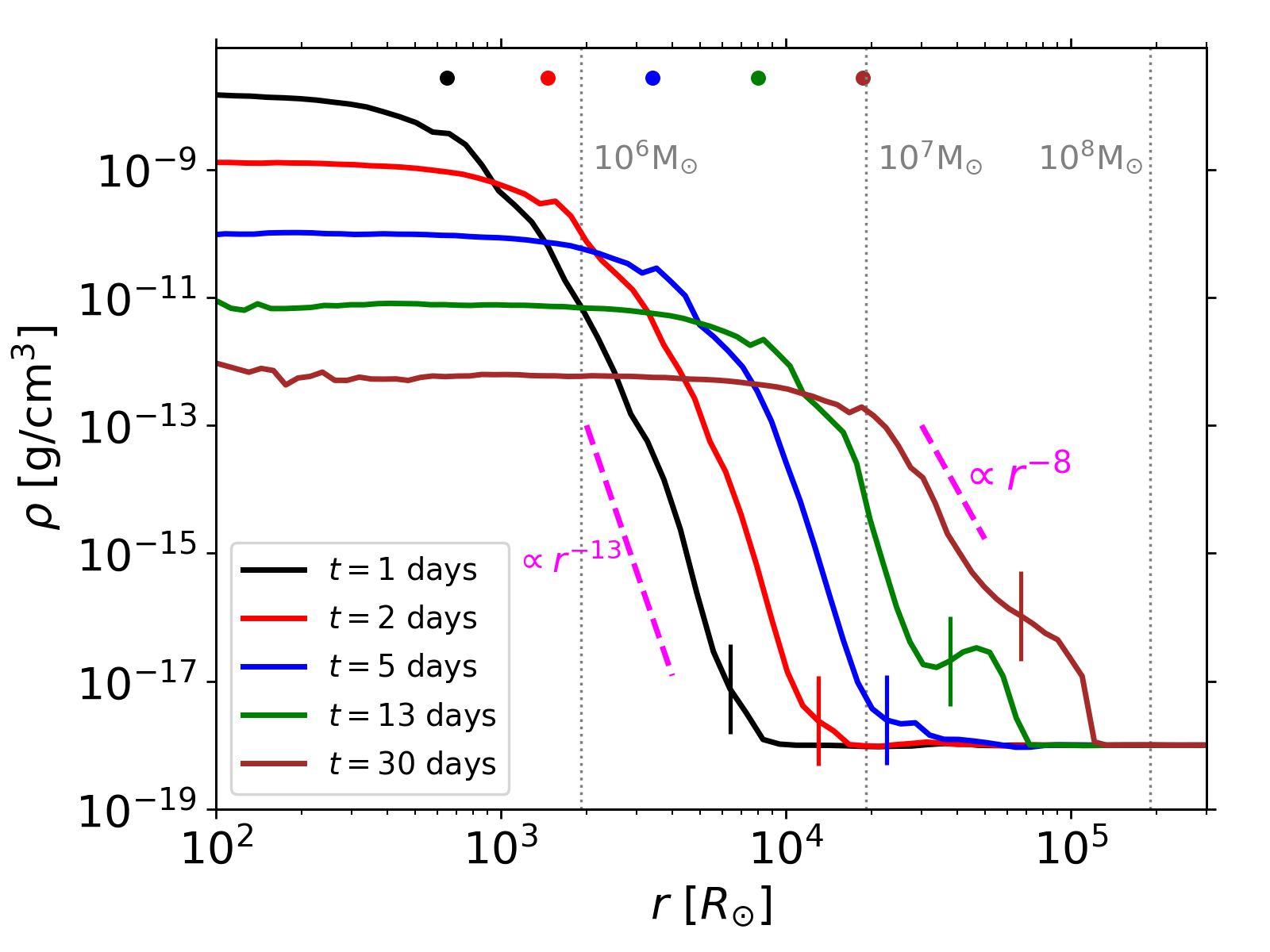}
    \includegraphics[width=8.6cm]{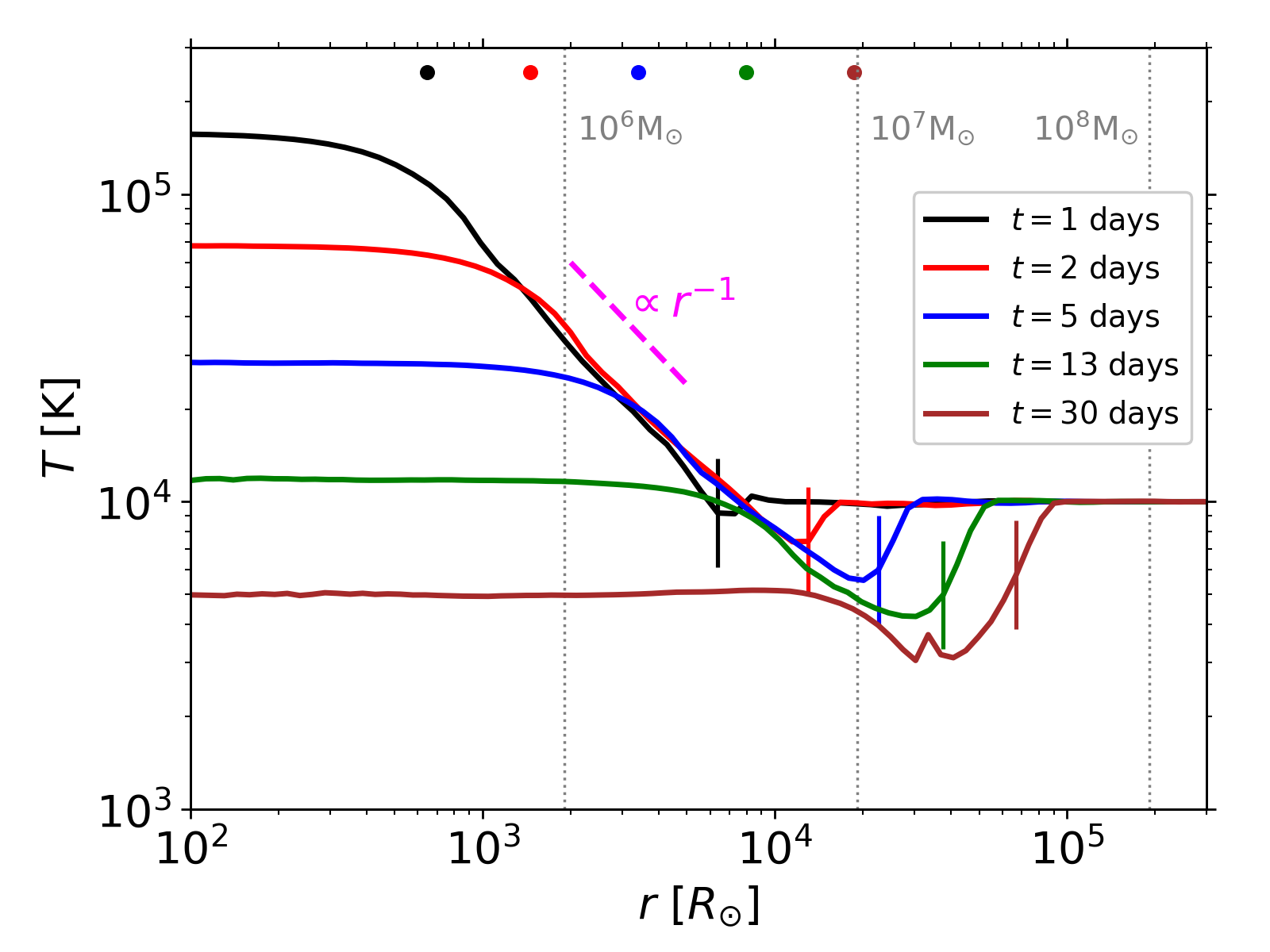}\\ 
    \includegraphics[width=8.6cm]{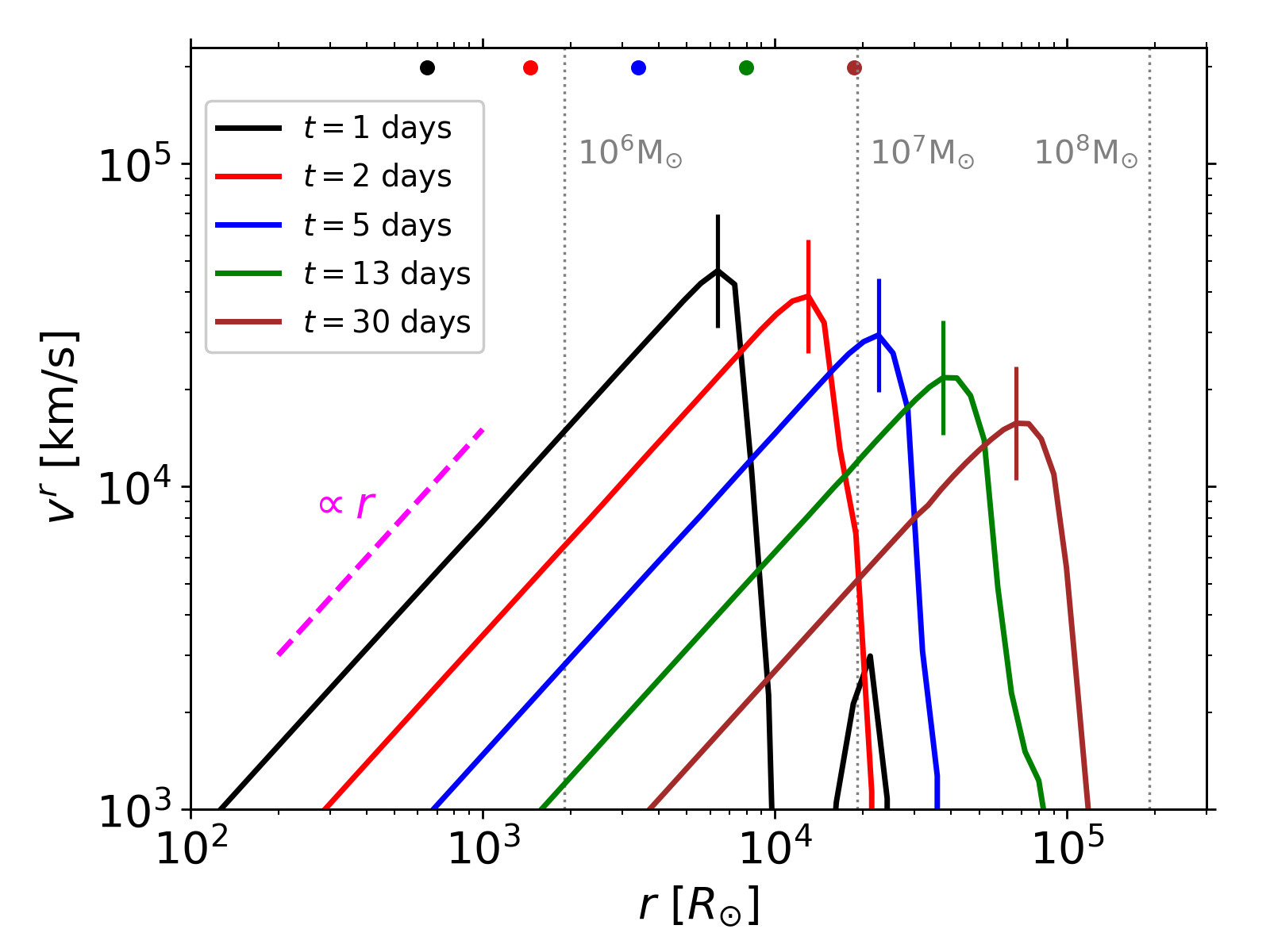}
    \includegraphics[width=8.6cm]{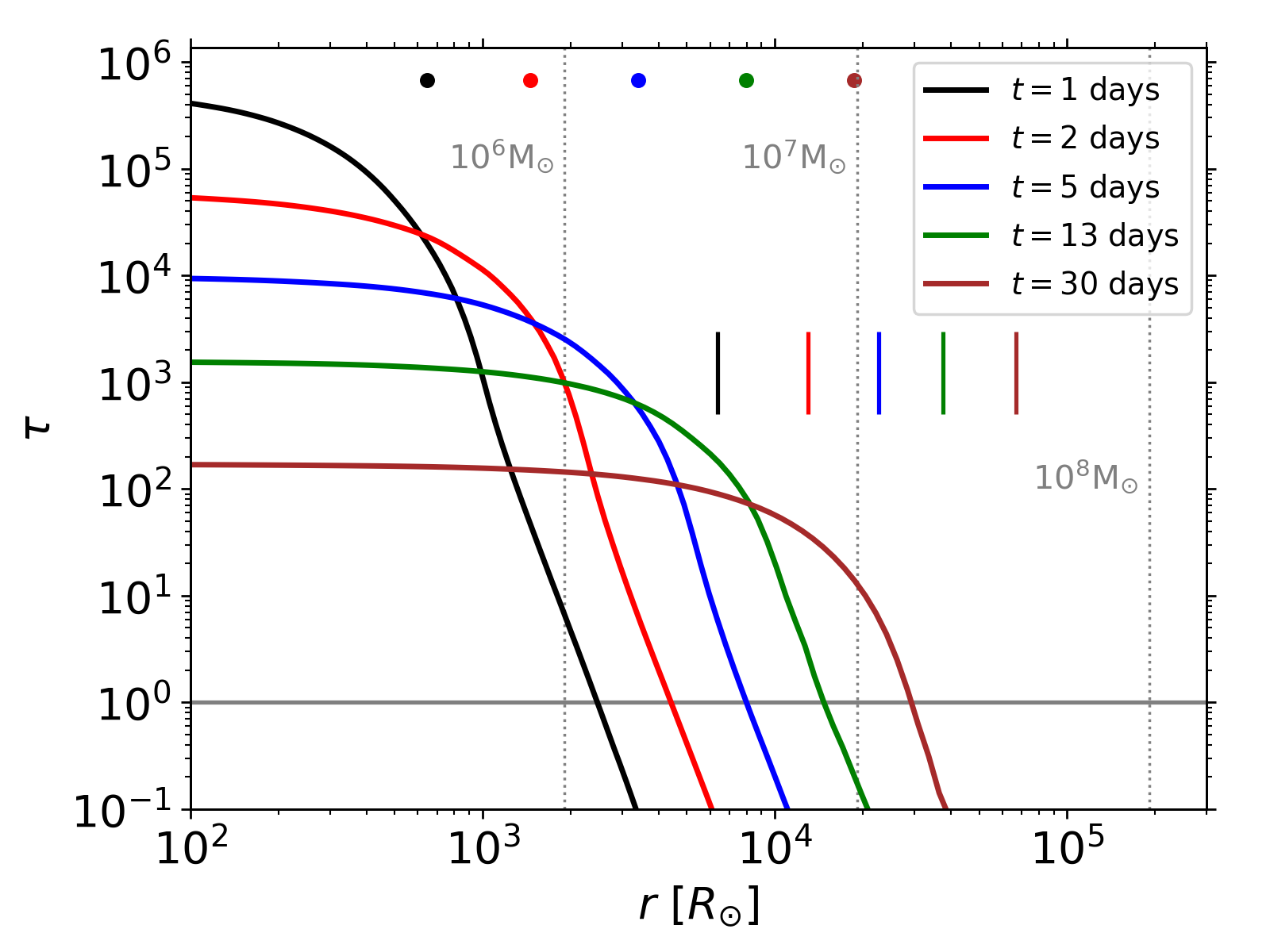}    
\caption{Spherically averaged density (\textit{top-left}), temperature (\textit{top-right}), radial velocity (\textit{bottom-left}) and optical depth (\textit{bottom-right}), of the expanding cloud produced in a collision between two $10\Rsol$ giants, as a function of radius from the collision point at five different times. The averages of the temperature and radial velocity are mass-weighted. The optical depth is estimated by radially integrating $\kappa\rho{\rm d}r$ inwards from $20\times$ the distance of the core from the collision point.  The three vertical lines show the distance from the BH with mass $10^{6}$, $10^{7}$, and $10^{8}\Msol$ at which the Keplerian velocity is the same as the initial relative velocity. In other words, by the time the outer edge reaches the distances, the gas would meet the central BH.  The dots indicate the locations of the cores at the five times, sharing the same colors with the lines. The vertical bars indicate the location of the outer edge where the expansion velocity is the maximum.}
	\label{fig:density_T_vr}
\end{figure*}

\begin{figure*}
	\centering
	\includegraphics[width=8.4cm]{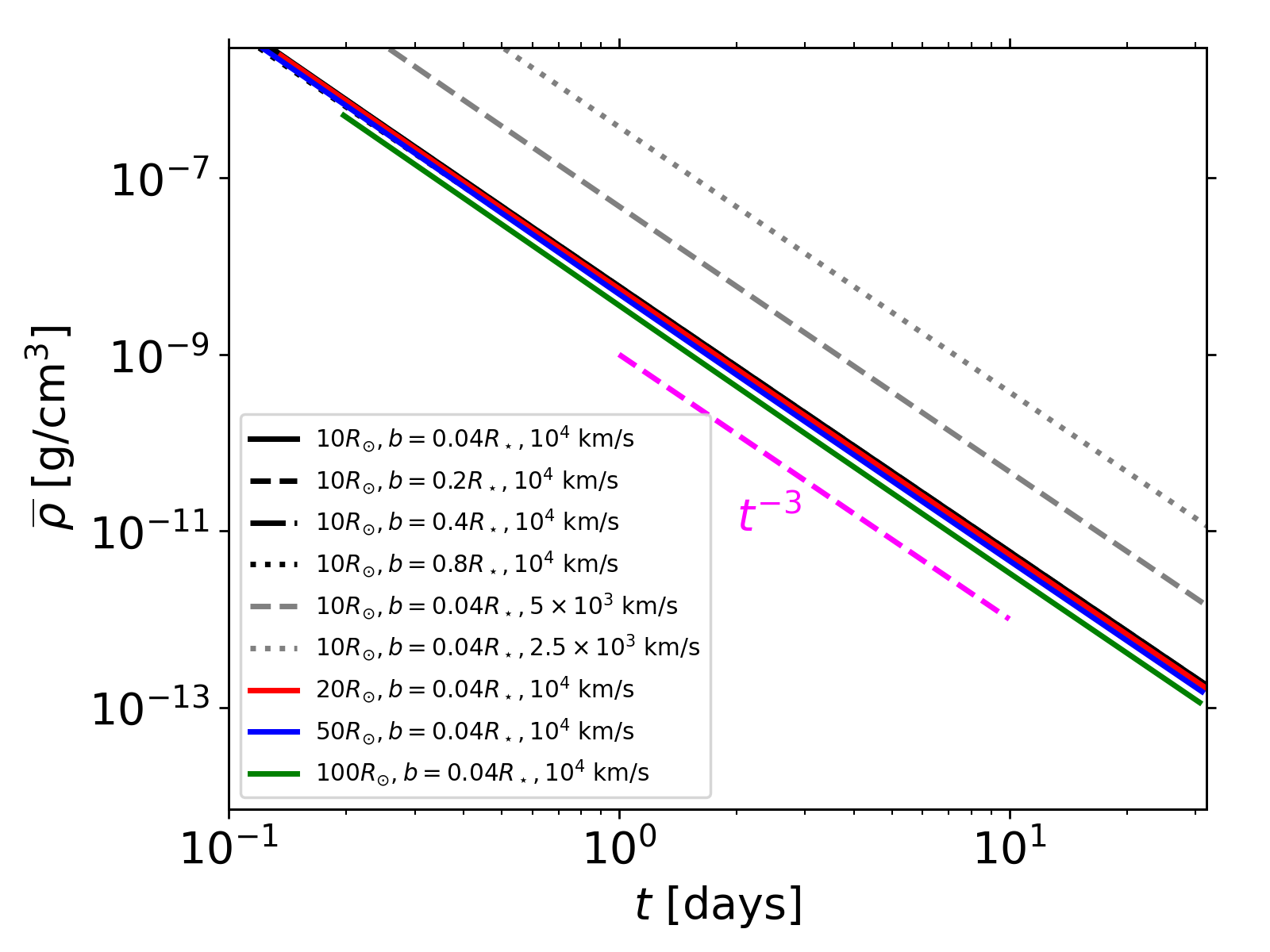}
	\includegraphics[width=8.4cm]{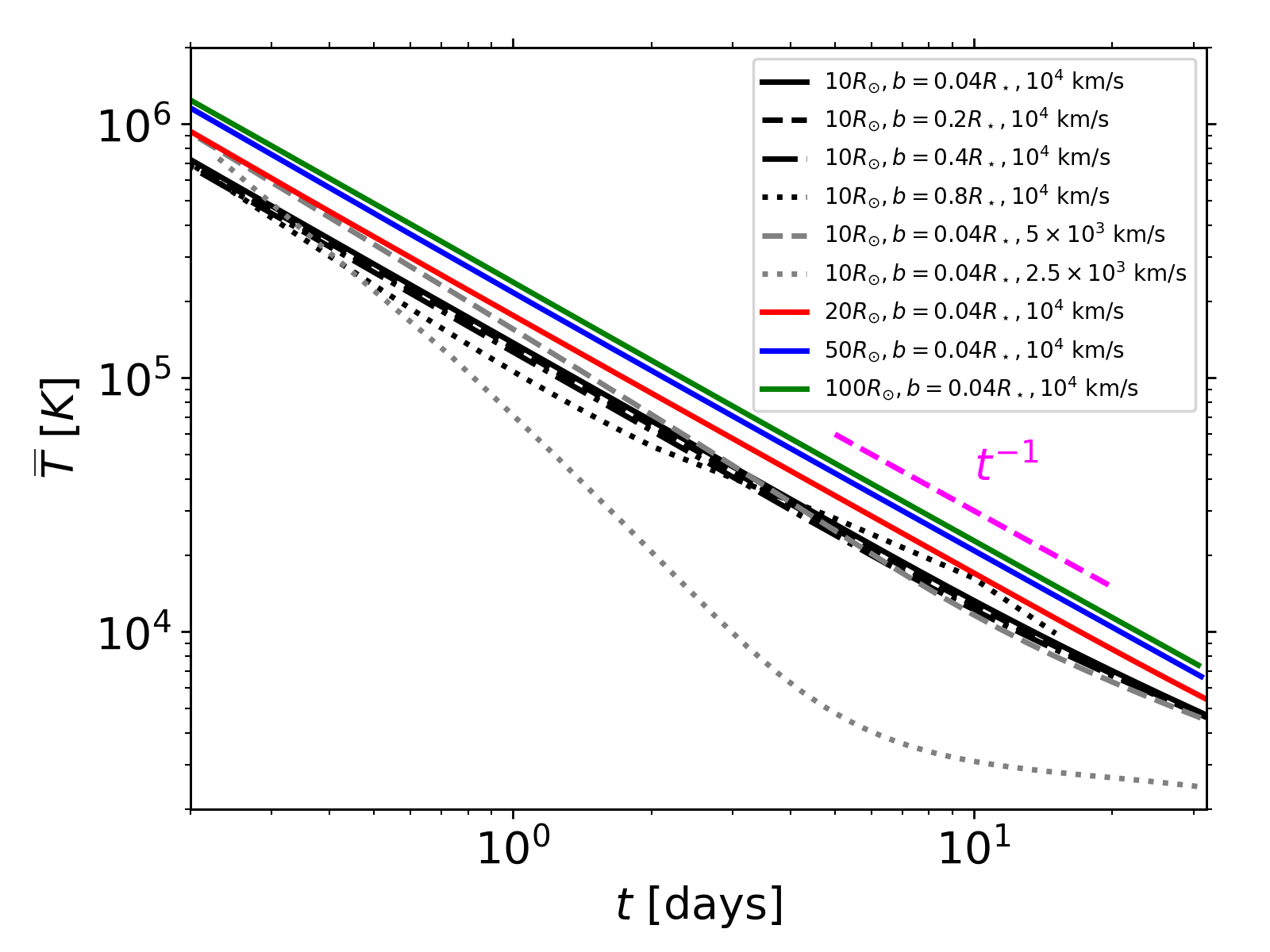}\\
    \hspace{0.1in}\includegraphics[width=8.2cm]{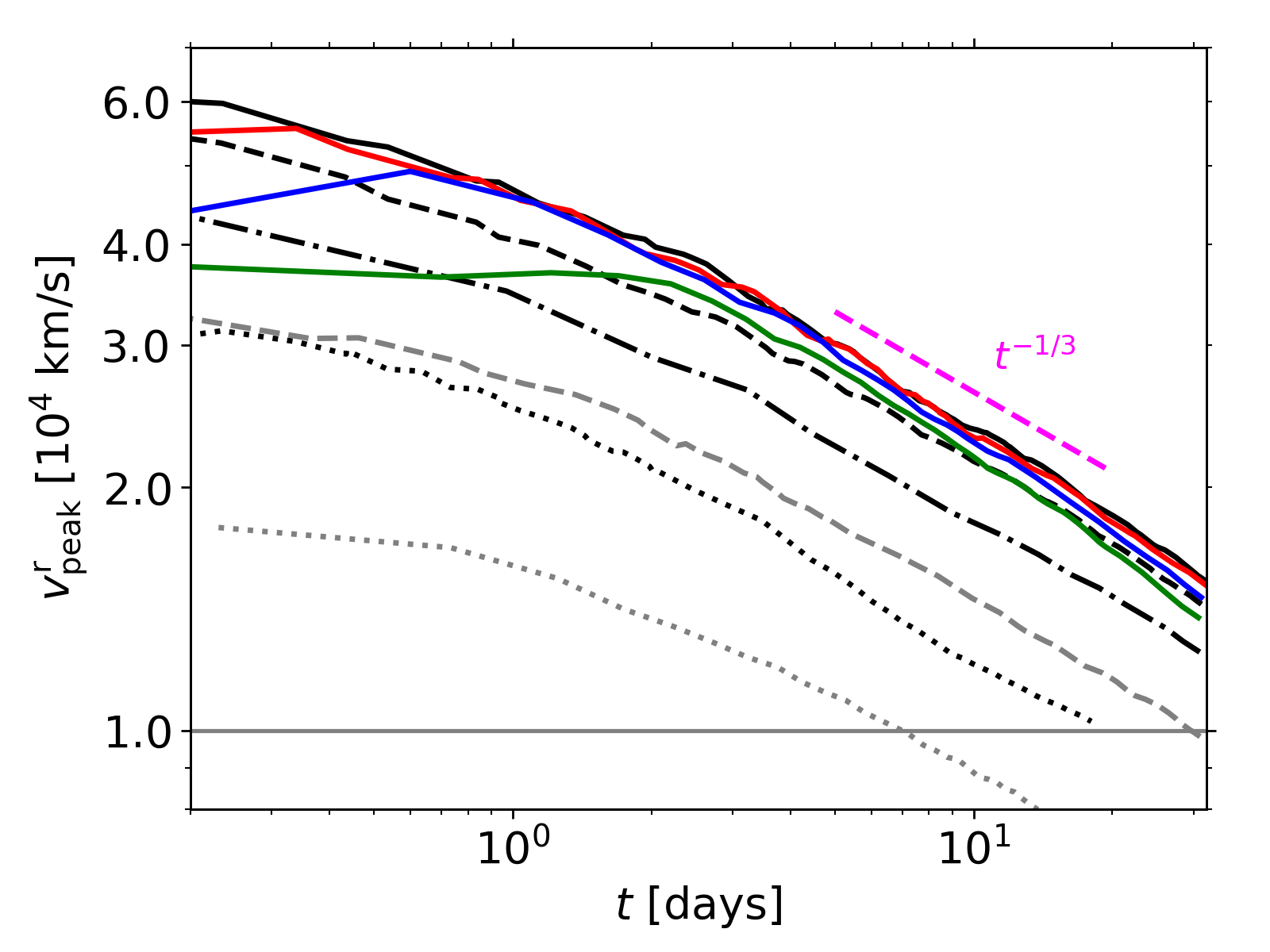}		
	\includegraphics[width=8.4cm]{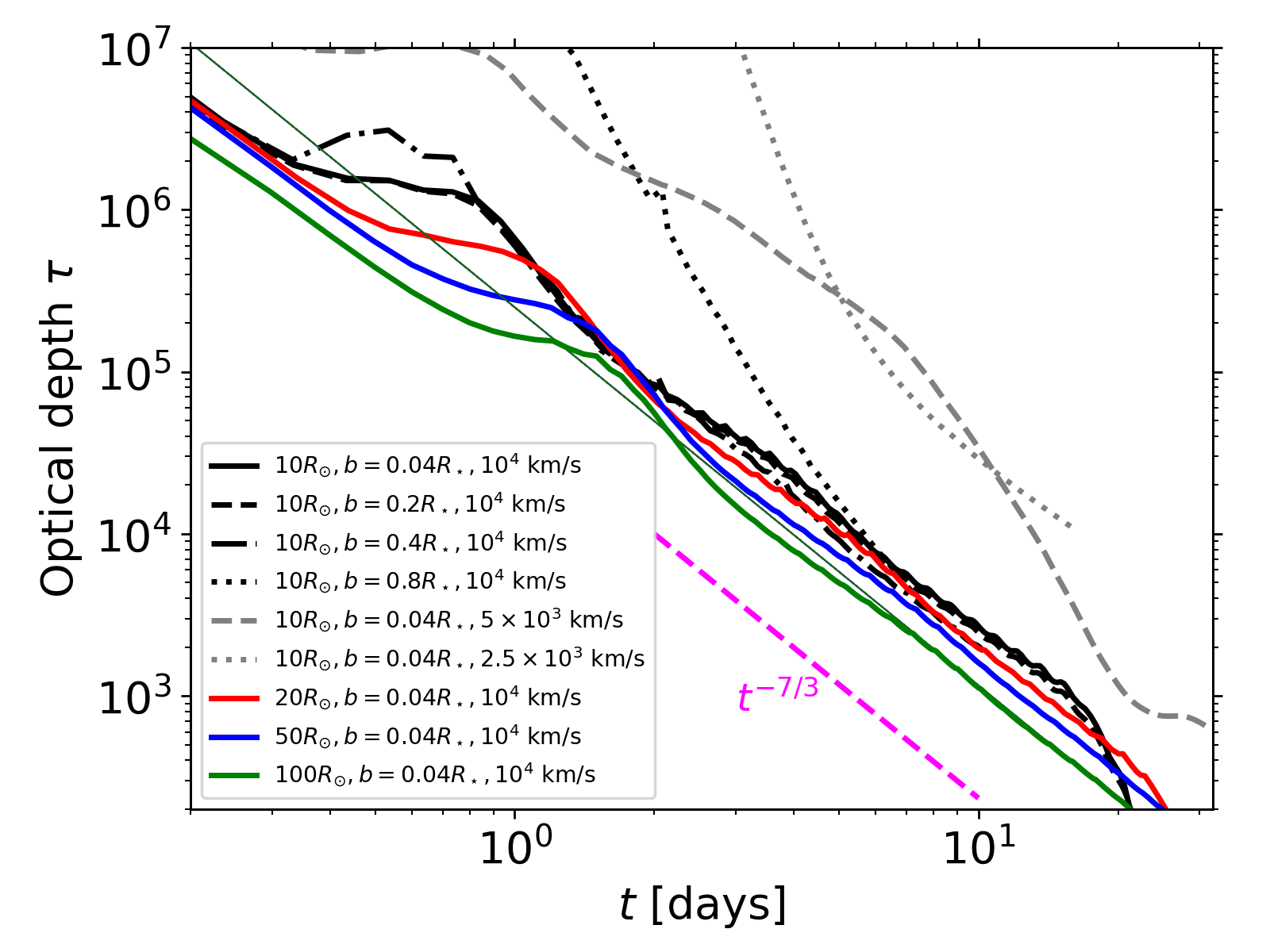}		
\caption{Average density $\overline{\rho}$ (\textit{top-left}), mass-weighted average of temperature $\overline{T}$ (\textit{top-right}), the peak expansion speed $v^{\rm r}_{\rm peak}$ (\textit{bottom-left}) and the surface average of the optical depth $\tau$ to the center(\textit{bottom-right}) of the cloud in all models, as a function of time since collision. The average density and temperature are estimated using the cells within a radius containing $75\%$ of the cloud mass. }
	\label{fig:density_vr2}
\end{figure*}

\begin{figure*}
	\centering
\includegraphics[width=17.5cm]{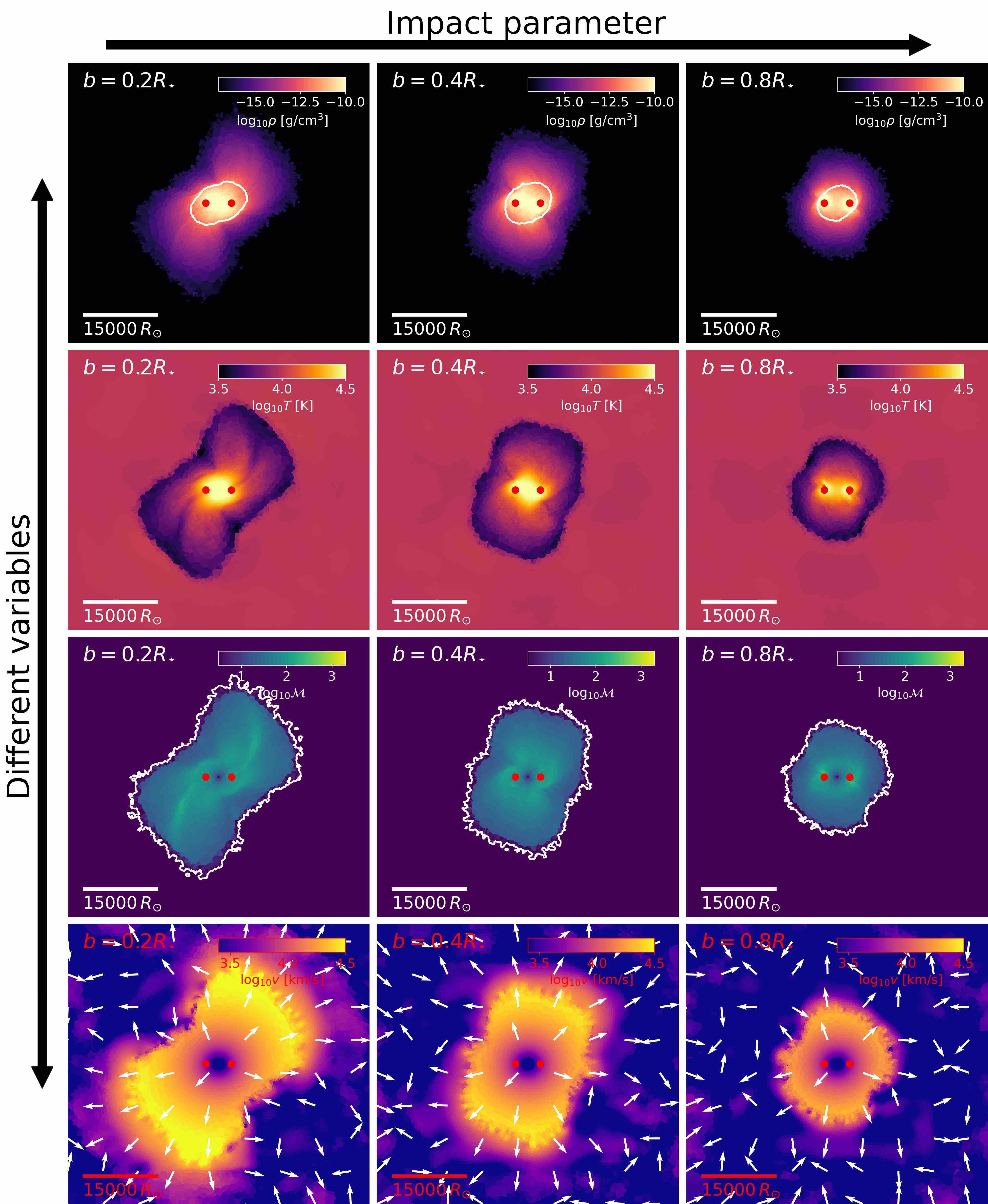}
\caption{Same as Figure~\ref{fig:evolution}, but for off-axis collisions ($b=0.2$, $0.4$, and $0.8\rstar$) between two giants with $\rstar=10\Rsol$ at $t\simeq 5$ days since collision.   }
	\label{fig:evolution3}
\end{figure*}

\section{Result}\label{sec:result}

\subsection{Overview}

We provide an overview of the evolution of the collision product using our fiducial model, e.g., head-on collision between the two $10\Rsol$ giants. We present in Figure~\ref{fig:evolution} (from \textit{top} to \textit{bottom}) the density $\rho$, the temperature $T$, the Mach number $\mathcal{M}$, and the speed in the mid-plane at four different times in our fiducial model.  

Initially, the two stars approach at $v_{\rm rel}\simeq 10^{4}$ km/s ($1^{\rm st}$ column). Since their first contact, the envelopes are continuously compressed due to the converging motion. Along the contact surface (the pronounced narrow feature across the center in the $2^{\rm nd}$ column, dubbed ``shock surface''), pressure gradients are built up and the temperature is raised above $10^{7}$ K due to adiabatic compression. As the later incoming gas collides supersonically with the pressure wall, shocks are created. Some of the very hot gas in the shock surface escapes radially perpendicular to the collision axis (or along the shock surface) with an opening angle of $\simeq 30^{\circ}$ and speeds of a few thousands km/s, which is not particularly high compared to the rest. At the strongest compression, a significant fraction of the kinetic energy is converted into heat energy ($\gtrsim 30\%$), which is already a few orders of magnitude greater than the total binding energy of the stars. When the pressure gradient exceeds the ram pressure, the compressed gas bounces off and expands quasi-spherically and homologously at supersonic speeds (see $3^{\rm rd}$ and $4^{\rm th}$ column panels in Figure~\ref{fig:evolution}). On top of the expanding motion, the converted heat energy continuously drives the outer part of the gas cloud to expand by the PdV work, meaning that some of the heat energy is converted back into kinetic energy. At the same time, the outer edge of the cloud supersonically collides with the background medium. This has two effects. First, mass piles up at the boundary between the gas cloud and the background medium, reducing the kinetic energy of the expansion front. Second, shocks are created, which dissipates the kinetic energy of the expansion front to heat energy. As a result of both effects, the expansion front slows down.

\subsection{Evolution of expanding cloud - parameter dependence}

\subsubsection{Fiducial case}
To describe the evolution of the expanding gas more quantitatively we show in Figure~\ref{fig:density_T_vr} the spherically-averaged density $\rho$ and (mass-weighted) temperature $T$, the expansion speed $v^{\rm r}$, and the area-weighted average of the optical depth $\tau$ over the solid angle for our fiducial model as a function of distance from the collision point at five logarithmically sampled times between $1$ and 30 days after collision. The density $\rho$ (\textit{top-left}) and the temperature $T$ (\textit{top-right}) of the inner regions of the expanding gas cloud are nearly constant. As the cloud expands adiabatically, the overall level of $\rho$ and $T$ drops while maintaining its slope: $\rho\simeq 10^{-8}\gram/\cm^{3}$ at $t\simeq 1$ day to $10^{-12}\gram/\cm^{3}$ at $t\simeq 30$ days, and $T\simeq 2\times 10^{5}$ K at $t=1$ day to $5\times 10^{3}$ K at $t\simeq 30$ days, at which point the cloud is cooler than the background medium. $\rho$ and $T$ outside the flat region decay towards the outer edge with a different steepness: the density drops following a power law of $\propto r^{-\lambda}$ with $\lambda\simeq 12-13 $ upon collision, gradually decreasing to $\lambda \simeq 8$ at $t\simeq 30$ days. But the temperature decays rather like $\propto r^{-1}$ at $1\lesssim t \lesssim 30$ days. The decaying slopes of $\rho$ and $T$ depend on $\rstar$, $b$, and $v_{\rm rel}$, but the dependence of the slope of $T$ is generally stronger. d$\ln\rho$/d$\ln r$ is almost the same, independent of $\rstar$ whereas -d$\ln T$/d$\ln r$ tends to be larger for larger $\rstar$ (e.g., $\lambda\simeq 2-3$ for $\rstar=100\Rsol$). d$\ln T$/d$\ln r$ is steeper for larger $b$ (e.g., $\lambda\simeq 2-3$ for $b=0.8\rstar$), while d$\ln\rho$/d$\ln r$ is only slightly less steeper for larger $b$ (e.g., $\lambda\simeq 12$ for $b=0.8\rstar$). The dependence of the slopes on $v_{\rm rel}$ is relatively weak: $\lambda$ for $\rho$ is almost same for $2500{\rm km/s}\leq v_{\rm rel}\leq 10^{4}{\rm km/s}$ and $\lambda$ for $T$ is slightly larger for smaller $v_{\rm rel}$ (e.g., $\lambda \simeq 1-1.5$ for $v_{\rm rel}=2500$ km/s).

As shown in the \textit{bottom-left} panel of Figure~\ref{fig:density_T_vr}, the cloud expands homologously, i.e., $v^{\rm r}\propto r$ or constant $v^{\rm r}$ at the same mass coordinate, which is also found in all other models. Right after the collision, the maximum expansion velocity at the outer edge is greater than the initial relative velocity by a factor of $\simeq 5$ and stays constant. The period of time with a constant peak $v^{\rm r}$ is very brief for this particular model ($\lesssim 0.1$ days). However the constant maximum $v^{\rm r}$ phase is longer for collisions with larger $\rstar$, which is illustrated in the \textit{bottom-right} panel of Figure~\ref{fig:density_vr2}. After the constant maximum $v^{\rm r}$ phase, the peak expansion velocity continuously decreases due to the interactions with the background medium. 

The gas cloud is initially optically thick. The optical depth to the center, estimated using an OPAL opacity table for Solar metallicity \citep{OPAL}, is $\tau\gtrsim 10^{5}$ at $t\simeq 1$ day, as demonstrated in the \textit{bottom-right} panel of Figure~\ref{fig:density_T_vr}. As it expands and cools, $\tau$ decreases following a power-law of $t^{-7/3}$ (see the \textit{bottom-right} panel of Figure~\ref{fig:density_vr2}), indicating that the entire cloud will become optically thin within 7 - 8 months, consistent with the analytic estimate by \citet{AmaroSeoane2023}. The nearly flat $\tau$ inside the cloud indicates that the transition from optically thick to completely optically thin may be prompt. 

\subsubsection{Comparison between models}

To further demonstrate the dependence of the stellar radius $R_{\star}$, the impact parameter $b$ and the initial relative velocity $v_{\rm rel}$, we compare in Figure~\ref{fig:density_vr2} the evolution of the same four quantities, shown in Figure~\ref{fig:density_T_vr} between different models. For a proper comparison, we estimate $\overline{\rho}$ as the average volume within a distance enclosing 75\% of the gas mass\footnote{Note that the radius enclosing 75\% of the cloud mass corresponds to the radius inside which $\rho$ and $T$ are constant, coinciding with the distance of the cores from the collision point.} and $\overline{T}$ as the mass-weighted average of $T$ within the same volume. As shown in the \textit{top} panels, $\overline{\rho}$ and $\overline{T}$ decrease over time, following a power-law of $t^{-3}$, and $t^{-1}$, respectively, almost independently of $R_{\star}$ and $b$ except for $\overline{T}$ with $v_{\rm rel}=2.5\times 10^{3}$ km/s. The $t^{-3}$ power-law for $\overline{\rho}$ is expected from an homologous expansion: $\rho\propto (v^{\rm r} t)^{-3}\propto t^{-3}$. As the $t^{-1}$-scaling relation for $\overline{T}$ suggests, the total (radiation + gas) internal energy at a given mass coordinate decreases like $t^{-1}$\footnote{Total specific energy $= 4\sigma T^{4}/[3c\rho] + k_{\rm B}T/[\mu m_{\rm p}]\propto t^{-1}$ because $T\propto t^{-1}$ and $\rho\propto t^{-3}$.}. The significant deviation from the $t^{-1}$ power-law for $v_{\rm rel}=2500$ km/s indicates that there is continuous energy exchange between gas at different mass shells. Unlike other cases where the radiation energy is dominant, in this case, the gas internal energy is comparable to the radiation energy and the total internal energy drops like $\propto t^{-4/3}$, resulting in a non-power law decay curve for $\overline{T}$. Although each of the two quantities, $\overline{\rho}$  and $\overline{T}$, tends to follow a single power-law, the degree to which their magnitudes depend on $R_{\star}$, $b$, and $v_{\rm rel}$ is different. $\overline{\rho}$ has a very weak dependence on $b$ and $\rstar$. $\overline{T}$ is insensitive to $b$ and weakly depends on $\rstar$: only a factor of 1.5 greater for $\rstar=100\Rsol$ than that $\rstar=10\Rsol$. 

 $v^{\rm r}_{\rm peak}$ stays constant upon collision at $(3-6)\times v_{\rm rel}$. The constant-$v^{\rm r}_{\rm peak}$  phase lasts longer for the case involved with stronger shocks (e.g., larger $\rstar$ for given $b$ and $v_{\rm rel}$). Eventually, $v^{\rm r}_{\rm peak}$ decreases over time because of the interactions with the background medium, following a power-law of $t^{-1/3}$ for all models. In particular, the peak expansion speed with varying $\rstar$ tends to asymptote to a single value at later times. As $b$ and  $v_{\rm rel}$ decrease, $v^{\rm r}_{\rm peak}$ is smaller at a given time. But the difference is at most by a factor of 3 for the collision parameters considered. 
 
 As explained for our fiducial model above, the optical depth is initially high at collision, $\tau> O(10^{6})$. The optical depth for most cases gradually decreases as the gas cloud expands, following a power-law $t^{-7/3}$, which is expected from the scaling relations of $\rho$ and $v^{\rm r}_{\rm peak}$: $\tau\propto \rho R_{\rm peak}\propto t^{-3} t^{2/3}\propto t^{-7/3}$ where $R_{\rm peak}$ is the location of the peak expansion speed $\simeq v^{\rm r}_{\rm peak}t\propto t^{2/3}$. Note that we assume a constant opacity to find the scaling relation given that the electron scattering is the dominant source of opacities in the gas cloud. The deviation from the $t^{-7/3}$ power-law relation becomes more significant as the collisions happen at lower $v_{\rm rel}$ and higher $b$.

\subsubsection{Fitting formulae}
 
 Combining all the scaling relations, we find that the average density $\overline{\rho}(t)$, mass-weighted average of temperature $\overline{T}$ peak expansion velocity $v^{\rm r}_{\rm peak}(t)$, size of the outer edge $R_{\rm peak}(t)$ and radial expansion speed $v^{\rm r}_{\rm peak}(r,t)$ after $t>5$ days can be well-described by the following analytic expressions,
\begin{align}\label{eq:rho}
    \overline{\rho}(t) &= 6\times10^{-10}{\rm g/cm}^{3} \left(\frac{t}{1 {\rm day}}\right)^{-3} \left(\frac{v_{\rm rel}}{10^{4}{\rm km/s}}\right)^{-3},\\
    \label{eq:Tave}
    \overline{T}(t) &=1.5\times10^{5} K  \left(\frac{t}{1 {\rm day}}\right)^{-1}\tan^{-1}(\sqrt{\rstar/10\Rsol}) \nonumber\\
    &\hspace{0.2in}\mathrm{ for} ~v_{\rm rel}\geq5000~ \mathrm{ km/s}~ \mathrm{and} ~b\lesssim 0.4\rstar ,\\
    \label{eq:tau}
    \tau(t) &= 2.5\times10^{5}  \left(\frac{t}{1 {\rm day}}\right)^{-7/3},\nonumber\\
    &\hspace{0.2in}\mathrm{ for} ~v_{\rm rel}>5000~ \mathrm{ km/s}~ \mathrm{and} ~b\lesssim 0.4\rstar ,\\
    \label{eq:vrpeak}
    v^{\rm r}_{\rm peak}(t) &= 50000 {\rm km/s} \left(\frac{t}{1 {\rm day}}\right)^{-1/3} \left(\frac{v_{\rm rel} }{10^{4}{\rm km/s}}\right)^{0.7} \left(\frac{b/R_{\star} + 5}{5}\right)^{-4},\\
    \label{eq:rpeak}
    R_{\rm peak}(t) &=6.5\times10^{14} \left(\frac{t}{1 {\rm day}}\right)^{2/3} \left(\frac{v_{\rm rel} }{10^{4}{\rm km/s}}\right)^{0.7} \left(\frac{b/R_{\star} + 5}{5}\right)^{-4},\\
    \label{eq:vr}
    v^{\rm r}(r,t) &= \begin{cases}
    v^{\rm r}_{\rm peak} \left(\frac{r}{R_{\rm peak}}\right)\hspace{0.1in} {\rm for } ~r\leq R_{\rm peak},\\
    0 \hspace{0.68in}{\rm for } ~r> R_{\rm peak},
    \end{cases}
\end{align}
where the expression for $R_{\rm peak}$ is found by analytically integrating $v^{\rm r}(r,t)$ over time. Note that $\overline{\rho}$ decays faster than that expected from the expression $3M_{\rm gas}/(4 \pi R_{\rm peak}^{3})\simeq t^{-2}$ because $\overline{\rho}$ follows the homologous relation whereas the peak expansion speed slows down so the outer edge expands slower than that expected for homologous expansion. 

Note that we do not include the term describing the dependence on $\rstar$ in most of the expressions above because of their very weak $\rstar$-dependence. On the other hand, the omission of the $v_{\rm rel}$-dependence in Equation~\ref{eq:Tave} for $\overline{T}$ is because of too small number of models with varying $v_{\rm rel}$ for reliable fitting. Instead, we have specified the range of $v_{\rm rel}$ where the equation is valid.

\subsection{Stellar core}

The cores move almost synchronously with the bulk of the gas. The orbit of the cores are barely affected by the collision: they remain unbound after collision and move away from each other at a speed almost same as the incoming speed. The distances from the collision point in our fiducial model at five different times are marked with circles in Figure~\ref{fig:density_T_vr}.

The mass bound to the cores is larger for larger $v_{\rm rel}$ and smaller $b$. But it is overall insignificant. For $b\leq 0.2\Rsol$ and $v_{\rm rel}\geq 5000$ km/s, the bound mass is less than $6\times 10^{-6}\Msol$. It is $\simeq 2\times10^{-3}\Msol$ for the model with $v_{\rm rel}=2500$ km/s and that with $b=0.4\Rsol$ and $\simeq 3\times10^{-2}\Msol$ for the model with $b=0.8\Rsol$.

\subsection{Conversion factor}

In this section, we investigate how much heat energy is created in collisions, which is closely related to the amount of energy that can be radiated away and potentially observed. We first define the conversion factor $\eta_{\rm rad}$ as the ratio of the total radiation energy to the initial kinetic energy,
\begin{align}\label{eq:eta}
    \eta_{\rm rad}(t) =\frac{\int aT(t)^{4}dV }{\int 0.5 \rho(t=0) v(t=0)^{2}dV},
\end{align}
where $a$ is the radiation constant and $dV$ is the volume element of each cell. Using $\eta_{\rm rad}$ one can estimate the total radiation energy as $\simeq  0.25\eta_{\rm rad} \mstar v_{\rm rel}^{2}$ for equal-mass collisions.  To distinguish gas that initially belonged to the stars from the background gas, we employ a selection condition using a passive scalar. The passive scalar is an artificial scalar quantity initially assigned to each cell which then evolves via advection without affecting the evolution of hydrodynamics quantities. The initial values of the passive scalar of the cells in the stars are one and that of the background cells is zero. So depending on the mass exchange (or mixing) between the cells, the passive scalar varies between zero (vacuum cells) and one (cells originally in the stars). We perform the integration over cells with the passive scalar $\gtrsim 0.1$. The value of $\eta_{\rm rad}$ is largely unaffected by the choice of the threshold of the passive scalar, provided that it is greater than 0. 

\begin{figure}
	\centering
	\includegraphics[width=8.6cm]{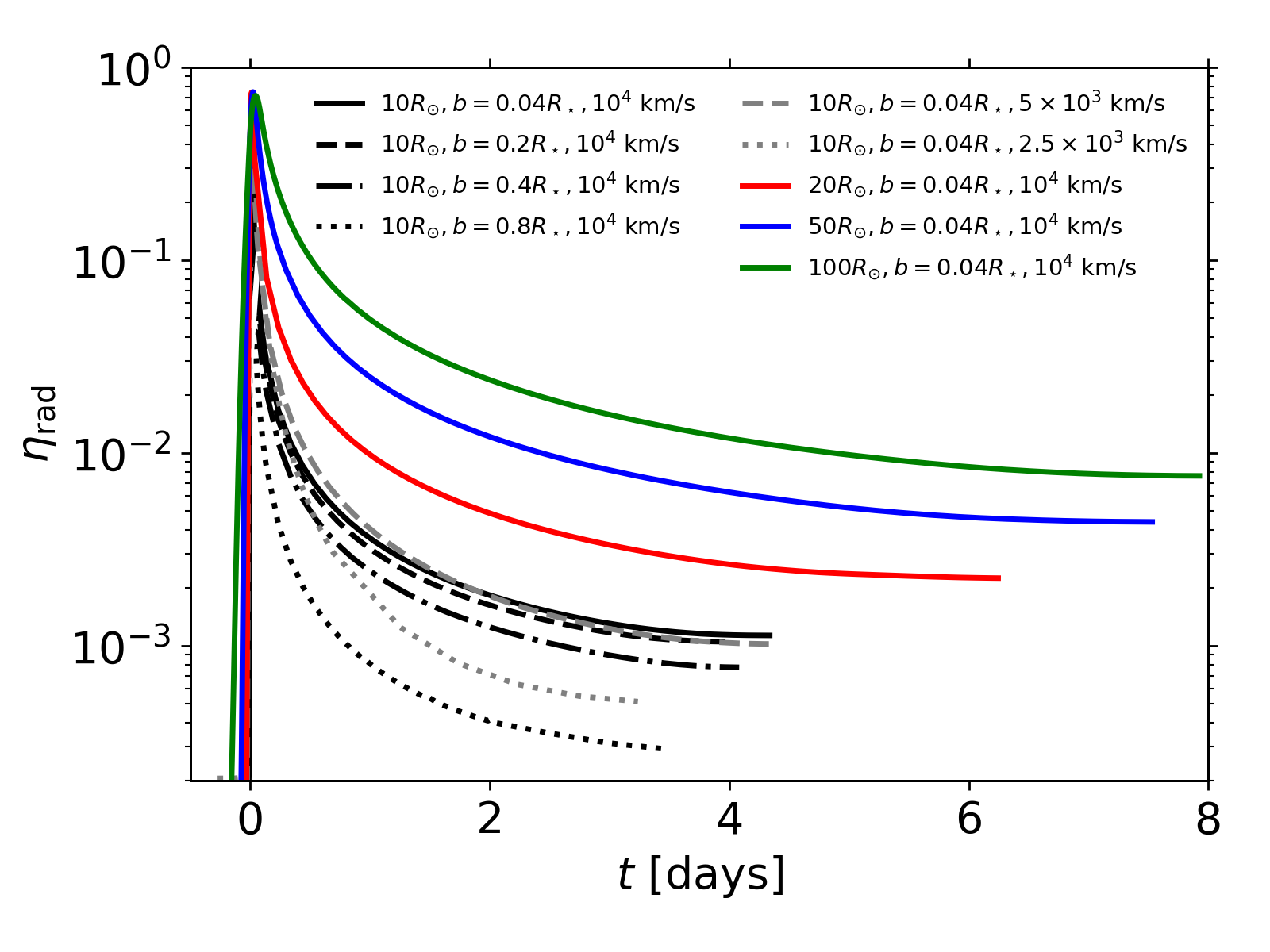}
\caption{The ratio of the radiation energy to the initial kinetic energy $\eta_{\rm rad}$ as a function of time, measured since collision, for all our models.}
	\label{fig:eta}
\end{figure}

We show $\eta_{\rm rad}$ for all our models in Figure~\ref{fig:eta} before the radiation energy in the optically thin gas becomes dominant. It is generally found that $\eta_{\rm rad}$ dramatically increases at collision to $\eta_{\rm rad}\simeq  0.1- 0.8$, meaning a significant fraction of the initial kinetic energy is converted into heat energy. The maximum conversion factors are summarized in Table~\ref{tab:LT}. Then as the cloud expands and cools, $\eta_{\rm rad}$ decreases down to $\lesssim 10^{-2}$. 
We see three clear post-peak trends of $\eta_{\rm rad}$. First, $\eta_{\rm rad}$ is larger when larger stars collide. Additionally, $\eta_{\rm rad}$ is approximately $\propto R_{\star}$ at any given time: $(1-2)\simeq 10^{-3}$ for $\rstar=10\Rsol$, $\simeq (3-4)\times10^{-3}$ for $\rstar=20\Rsol$, $\simeq \times10^{-2}$ for $R_{\star}=50\Rsol$, and $\simeq 2\times 10^{-2}$ for $\rstar = 100\Rsol$ at $t\simeq 3$ days. We attribute this positive correlation between $\eta_{\rm rad}$ and $R_{\star}$ to the fact that for the same relative velocity, larger (cooler) stars collide at higher $\mathcal{M}$, resulting in stronger shocks over a wider contact surface ($\propto R_{\star}$). Second, $\eta_{\rm rad}$ is almost the same when $b\lesssim 0.2\rstar$, while $\eta_{\rm rad}$ begins to decrease with $b$ when $b\gtrsim 0.2\rstar$. This trend is somewhat expected given that as $b$ increases, the mass of gas that is shocked at collision decreases. Lastly, $\eta_{\rm rad}$ decreases with $v_{\rm rel}$ because of lower $\mathcal{M}$ collisions for given sound speed (i.e., the same star). $\eta_{\rm rad}$ at $v_{\rm rel}=5000$ km/s is almost the same as that at $v_{\rm rel}=10000$ km/s, but $\eta_{\rm rad}$ at $v_{\rm rel}=2500$ km/s is lower by a factor of $\simeq 2$ than that for our fiducial case. The overall levels of the conversion factors that we obtain are comparable to what \cite{AmaroSeoane2023} imposed in order for their analytical model to match with the observed object ZTF19acboexm (see their Figure 9).

We can also define the conversion factor for the ram pressure of gas moving at supersonic speeds,
\begin{align}\label{eq:eta_rem}
    \eta_{\rm ram}(t) =\frac{\int \rho(t) v(t)^{2} dV }{\int \rho(t=0) v(t=0)^{2}dV},
\end{align}
where the integration in the denominator is carried out over cells for which the passive scalar $>0.1$, and that in the numerator the integration is carried out only over cells with supersonic speeds, $\mathcal{M}\geq1$. As illustrated in the $3^{\rm rd}$ column panels of Figure~\ref{fig:evolution} and \ref{fig:evolution3}, almost all the gas is supersonically expanding. As a result, $1-\eta_{\rm ram}\simeq \eta_{\rm rad}$.

\begin{figure}
	\centering
\includegraphics[width=8.5cm]{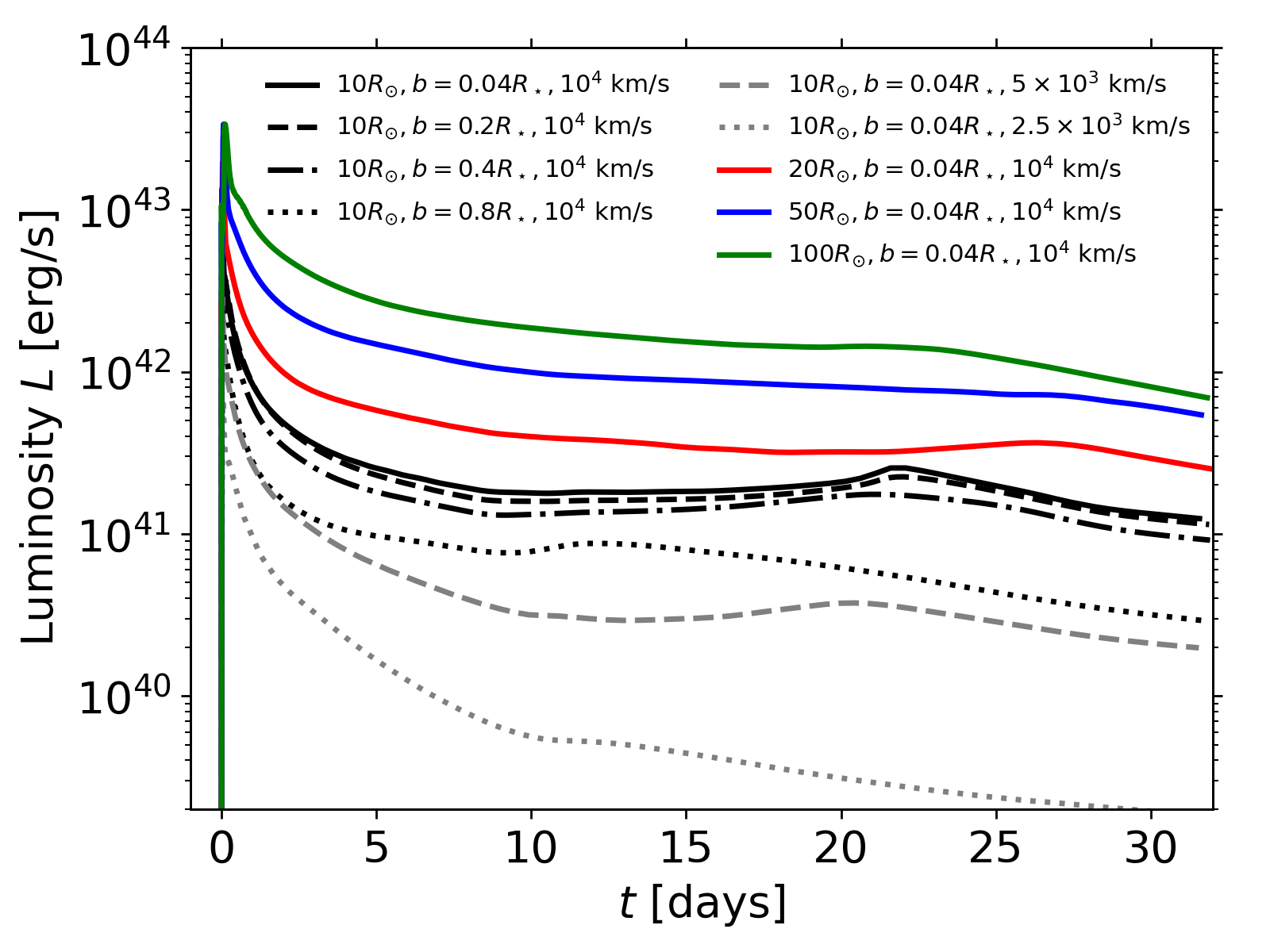}
\includegraphics[width=8.5cm]{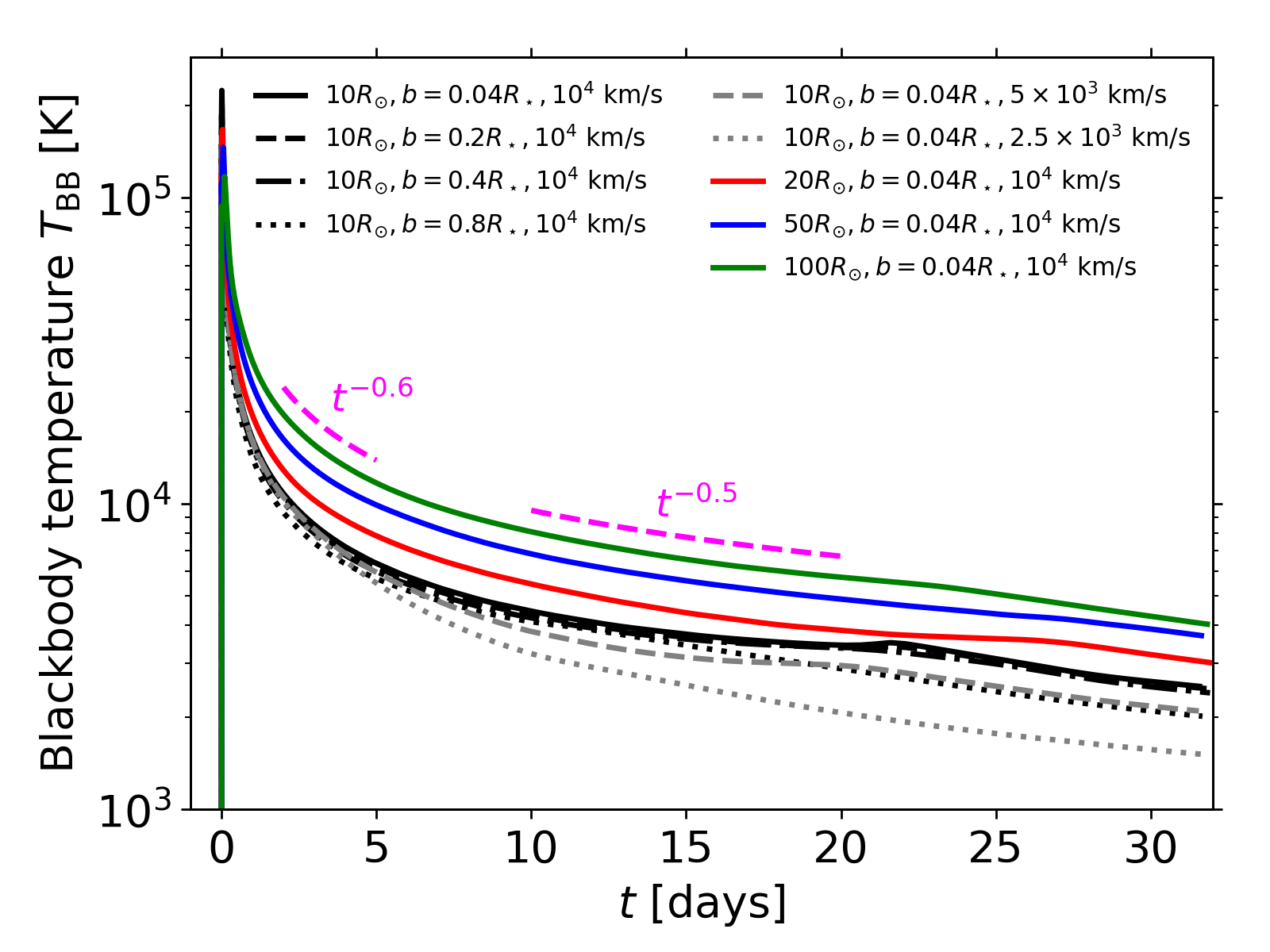}
\includegraphics[width=8.5cm]{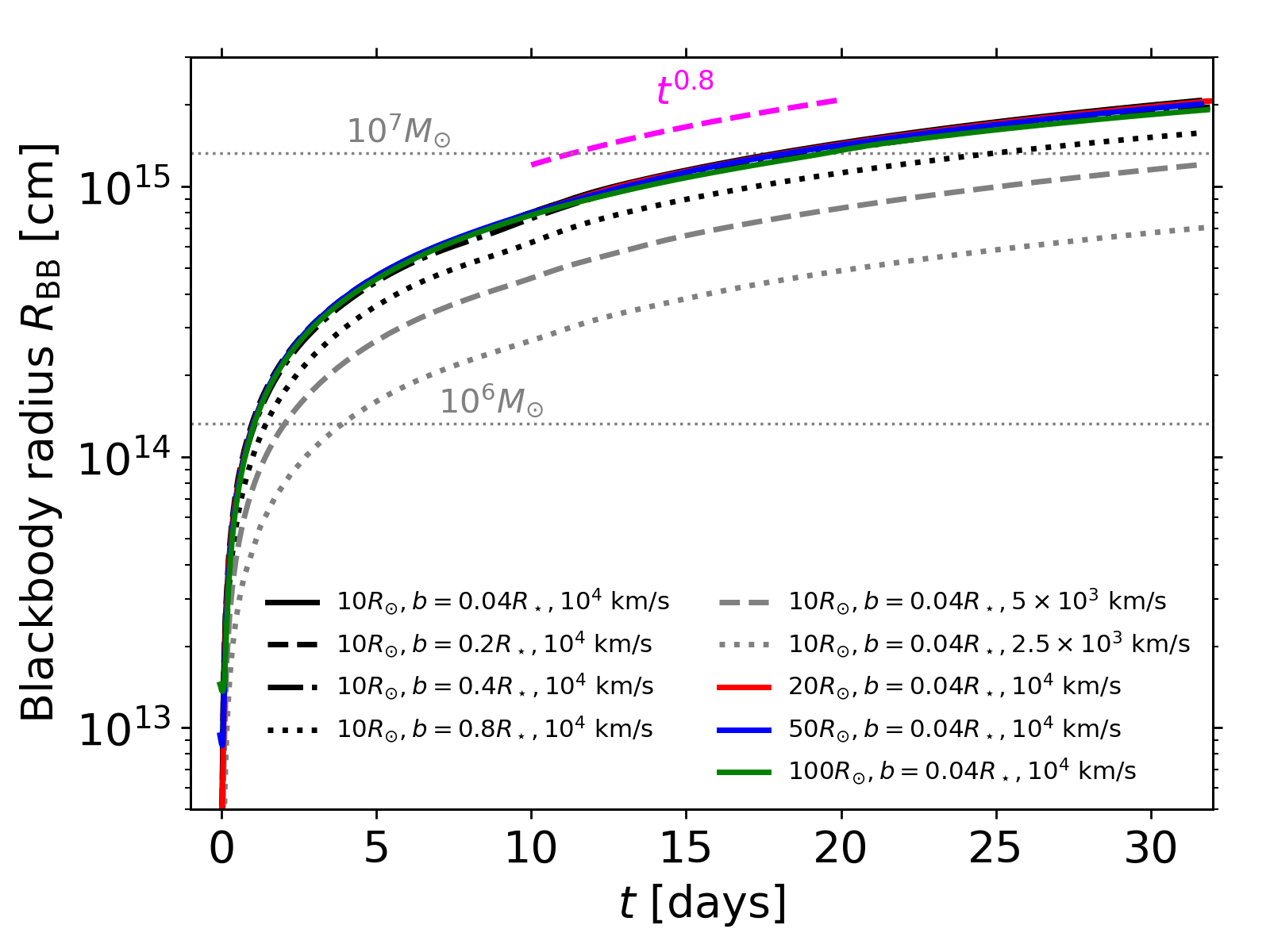}
\caption{Bolometric luminosity $L$ (\textit{top}), blackbody temperature $T_{\rm BB}$ (\textit{middle}), and blackbody radius $R_{\rm BB}$ (\textit{bottom}), estimated for stellar collisions using Equations ~\ref{eq:Abb} and \ref{eq:L1}. The dotted grey horizontal lines in the \textit{bottom} panel indicate the distances of the collision from the black hole with $M_{\rm BH}=10^{6}\Msol$ ($\simeq 10^{14}\cm$) and $10^{7}\Msol$ ($\simeq 10^{15}\cm$). The magenta guide lines show the power-law that describes the quantity shown in the last two panels. }
	\label{fig:observables}
\end{figure}

\subsection{Observables}\label{subsec:observables}
We estimate the luminosity $L$, blackbody radius $R_{\rm BB}$, and temperature $T_{\rm BB}$, using the radiation energy and the local cooling time $t_{\rm cool}$. We first construct a spherical grid with an extremely small opening polar angle ($\theta\simeq10^{-10}$ radians) to avoid the singularity at the poles, radially extending out to near the outer boundary of the domain. The grid in the radial direction is logarithmically divided, i.e., constant $\Delta r/r$ where $\Delta r$ is the cell size at $r$, while that in the $\theta$ and $\phi$ directions are linearly divided, i.e., constant $\Delta \theta$ and $\Delta \phi$. The number of grids in $r$, $\theta$, and $\phi$ are (800, 600, 600), which we confirmed to give converging estimates for the observables. We then identify the photosphere at which the optical depth $\tau\simeq 1$. $\tau$ is integrated along each $r$-path with the opacity found using an OPAL opacity table for Solar metallicity \citep{OPAL}. The photospheric area is,
\begin{align}\label{eq:Abb}
    A_{\rm BB} = \int_{0}^{2\pi}\int_{\simeq 0}^{\simeq\pi} r(\tau=1)^{2}\sin\theta dr d\theta d\phi,
\end{align}
which gives the effective size of the emitting region or blackbody radius $R_{\rm BB}=(A_{\rm BB}/4\pi)^{1/2}$.

We attempt to bracket the range of realistic radiated luminosity from the collision event by employing two different methods, each of which places different weights on the contribution from the gas cloud layers (the inner regions or outer regions near the photosphere) within the identified photosphere. Our estimates should be accurate at an order-of-magnitude level. 
However, for more accurate modeling of light curves, we will carry out detailed non-equilibrium radiation transport calculations in future follow-up work dedicated to estimating light curves and spectra. 

In both methods, the total luminosity for each radial path is estimated by summing the contributions from the cells with the local cooling time $t_{\rm cool}$ shorter than the evolution time $t$ within the photosphere.  Here, $t_{\rm cool}$ is defined as $h_{\rho}\tau(1+u_{\rm gas}/u_{\rm rad})/c$ where $h_{\rho}$ is the density moment scale height inside the photosphere and $u_{\rm rad}$ ($u_{\rm gas}$) the radiation (gas thermal) energy. However, the difference between the two methods is the assumption for how most of the radiation energy is radiated away. In one method, we assume that the total radiation energy within the photosphere is radiated away over a time comparable to the cooling time at the base of the cloud. Under this assumption, the inner regions tend to dominate the luminosity. We first integrate the total radiation energy along the radial path and divide it by the cooling time at the base of the cloud $t_{\rm cool,max}$, i.e., the longest cooling time which is no longer than $t$, or
\begin{align}\label{eq:L1}
    L_{1} = \int_{0}^{2\pi}\int_{\simeq 0}^{\simeq\pi} \left[\int_{r(t_{\rm cool}=t)}^{r(\tau=1)} aT^{4} r^{2}\sin\theta dr\right] t_{\rm cool, max}(\theta,\phi)^{-1} d\theta d\phi.
\end{align}
In the second, we assume that the radiation energy of each cell is radiated away over the local cooling time. So the total luminosity is estimated,
\begin{align}\label{eq:L2}
    L_{2}= \int_{0}^{2\pi}\int_{\simeq 0}^{\simeq\pi} \int_{r(t_{\rm cool}=t)}^{r(\tau=1)} aT^{4} t_{\rm cool}(r,\theta,\phi)^{-1} r^{2}\sin\theta dr d\theta d\phi.
\end{align}
In this method, the outer regions near the photosphere dominate the luminosity. As stressed before, the evolution of the hydrodynamics quantities for optically thin gas (i.e., outer region near the photosphere) in our simulations is intrinsically less accurate than those for optically thick gas. Hence, $L_{1}$ should be considered more consistent with our hydrodynamical scheme. We find
that the shapes of the $L_{1}$ and $L_{2}$ lightcurves are very similar. However, $L_{1}$ is consistently smaller than $L_{2}$ by a factor of $\simeq10$. For this reason, we present $L_{1}$ and the resulting blackbody temperature $T_{\rm BB,1}=(L_{1}/\sigma A_{\rm BB})^{1/4}$ in this section and those from Equation~\ref{eq:L2} in Appendix~\ref{appen1}.

Figure~\ref{fig:observables} shows $L_{1}$ (\textit{top}), $T_{\rm BB,1}$ (\textit{middle}), and $R_{\rm BB}$ (\textit{middle}) as a function of time measured since collision for all our models\footnote{While this paper is under review, we published \citet{Dessart+2023} in which we conducted detailed radiation transfer calculations for the observables of BDCs using the time-dependent radiation transfer code {\tt CMFGEN } \citep{Hillier+2012}. The more accurately estimated luminosity and temperature are in good agreement with our order-of-magnitude estimates assuming the inner regions dominating the luminosity (Equation~\ref{eq:L1}) shown in Figure~\ref{fig:observables}. }. Note that the luminosity and the blackbody temperature differ depending on the assumption of radiation (Equations~\ref{eq:L1} and \ref{eq:L2}), but $R_{\rm BB}$ is independent of the assumption. The luminosity increases dramatically to its peak at collision. The peak luminosity is $L_{1}\gtrsim 10^{41}-10^{43}$ erg/s ($L_{2}\gtrsim 10^{42}-10^{44}$), which is higher for larger $R_{\star}$ and smaller $b$ and higher $v_{\rm rel}$, which has the same trend as $\eta_{\rm rad}$. The temperature at peak is $T_{\rm BB, 1}\simeq 10^{5}$ K. Because $T_{\rm BB}\propto L^{1/4}$, $T_{\rm BB, 2}$ is greater than $T_{\rm BB, 1}$ by less than a factor of 2. We summarize peak $L$ and $T_{\rm BB}$ at peak for all our models in Table~\ref{tab:LT}. Subsequently both $L$ and $T_{\rm BB}$, independent of the assumption for the diffusion time (so both $L_{1}$ and $L_{2}$), decrease following a power-law $\propto t^{-\xi}$ with $\xi$ slightly differing at early and late times. $L$ at $t\lesssim 5$ days reveals a decaying curve with $\xi \simeq 0.7-0.8$, followed by a slower decay with $\xi \simeq 0.4$ at $t\gtrsim 5$ days. $L$ therefore decreases by a factor of 10 for the first 5 days. The decay in $L$ for the next 30 days is relatively small, by only by a factor of a few. The change in $\xi$ for $T_{\rm BB}$ is very mild: $\xi \simeq 0.6$ at $t\lesssim 5$ days and $\simeq 0.5$ at $t\gtrsim 5$ days. $T_{\rm BB}$ decreases from $\simeq (1-2)\times10^{5}$ K at collision to $10^{4}$ K at $5-15$ days, $(4-6)\times10^{3}$ K at $30$ days. This means the collision will be bright in extreme UV at collision which shifts to optical on a time scale of a month. Lastly, $R_{\rm BB}$ increases to $\simeq 10^{15}$cm in 30 days, approximately following power-law growth of $\propto t^{0.8}$.

The light curves from our simulations reveal some differences  from that analytically predicted by \citet{AmaroSeoane2023}. Assuming a constant $\eta$ comparable to the minimum $\eta_{\rm rad}$ shown in Figure~\ref{fig:eta}, their analytic model predicts a peak luminosity consistent with the numerically integrated peak luminosity shown in Figure~\ref{fig:observables}. However,  the luminosity from their analytic model peaks at a few days after collision and subsequently decays faster. We attribute these discrepancies to the difference in the way of calculating the luminosity. In their analytic model, the luminosity was estimated under the assumption that $\eta$ does not change over time and the total radiation energy within the gas cloud is radiated away instantaneously on a time scale comparable to the longest possible photon cooling time at any given time (e.g., based on the optical depth to the center). On the other hand, in this work, we take into account the time-dependent contributions (e.g., adiabatic loss of energy 
due to expansion) of the cloud. 

The observables estimated in this section are driven by stellar collisions. But given the fact that these collisions occur near a SMBH, the expanding gas cloud and the nearby BH would very likely interact, generating a possibly even brighter flare, which we discuss in \S~\ref{subsec:bhgas}.

\begin{table*}
\begin{tabular}{ c c c c | c c c c c c} 
\hline
Model number  &  $\rstar$  & $v_{\rm rel}$  & $b$ & $\eta_{\rm peak}$ & $L_{\rm peak,1}$ & $L_{\rm peak,2}$ & $T_{\rm BB,peak,1}$ & $T_{\rm BB,peak,2}$ \vspace{0.02in}\\
- &  $\Rsol$ & $10^{3}$~km/s & $\rstar$ & - & $10^{43}$erg/s  & $10^{43}$erg/s  & $10^{5}$K & $10^{5}$K\\
\hline
1 & 10 & 10 & 0.04  & 0.69 &  9.6 & 1.3 & 3.0 & 2.1 \\
2 & 10 & 10 & 0.2  & 0.59  &  9.3 & 1.3 & 2.9 & 2.0 \\
3 & 10 & 10 & 0.4  & 0.41  & 6.5 & 1.0 & 2.7 & 1.9 \\
4 & 10 & 10 & 0.8  & 0.13  &  4.2 & 0.7 & 2.7 & 1.7 \\
5 & 10 & 5 & 0.04  & 0.69  &   2.1 & 0.3 & 2.2 & 1.4 \\
6 & 10 & 2.5 & 0.04  & 0.56 &   0.5 & 0.1 & 1.0 & 1.0 \\
7 & 20 & 10 & 0.04  & 0.65 & 13 & 2.0 & 2.7 & 1.7 \\
8 & 50 & 10 & 0.04  & 0.73 & 26 & 3.4 & 2.2 & 1.4 \\
9 & 100 & 10 & 0.04 & 0.73 & 26 & 3.4 & 1.7 & 1.1 \\
\hline
\end{tabular}
\caption{Peak conversion factor $\eta$, luminosity at peak $L_{\rm peak}$ and blackbody temperature at peak $T_{\rm BB,peak}$ for each model, using Equation \ref{eq:L1} ($L_{\rm peak,1}$ and $T_{\rm BB,peak,1}$) and Equation \ref{eq:L2} ($L_{\rm peak,2}$ and $T_{\rm BB,peak,2}$).  }\label{tab:LT}
\end{table*}

\section{Discussion}\label{sec:discussion}

\subsection{Interaction of gas cloud with interstellar medium}\label{subsec:ism}

In this work, we simulated high-velocity collisions of giants surrounded by a medium with a constant density of $10^{-18}$g/cm$^{3}$ and temperature of $10^{4}$ K. As the cloud expands, it collides inelastically with the background medium, which results in the continuous decrease in the kinetic energy of the expansion front. In addition, the collision between the outer edge of the cloud and the background medium can create shocks, converting the kinetic energy into heat energy. The net effect is the deceleration of the gas cloud, deviating from an homologous behavior, which is also found from our simulations where the velocity of the outer edge decreases following $t^{-1/3}$. This impact of the surrounding medium would be faster if the colliding stars were initially embedded in a denser medium. For example, the rising slope of $\eta_{\rm rad}$ would be less steep for the case with lower-density background gas. Given the supersonic motion of the cloud, how the cloud expands would not be significantly affected by the temperature of the background medium for a given background density. However the evolution of $\eta_{\rm rad}$ would be changed depending on the background temperature. In fact, we performed extra simulations with different background temperatures ($100 -5000$ K), showing that while the expansion properties of the cloud (e.g., $\overline{\rho}$, $\overline{T}$, and $v_{\rm peak}^{\rm r}$) are almost independent of the background temperature, $\eta_{\rm rad}$ tends to be lower at the local minimum and increases more slowly afterward for a lower background temperature.

 Although the deviation from an homologous expansion was only found near the outer edge for the duration of our simulations, as an order-of-magnitude estimate, the motion of the entire gas cloud would become completely deviated from an homologous expansion when the swept-up mass is comparable to the mass of the cloud,
\begin{align}
    t_{\rm non-homologous} &\simeq \frac{3M_{\rm gas}}{4\pi\rho_{\rm ISM} (v^{\rm r})^{3}},\nonumber\\
             & \simeq 1100~{\rm days}\left(\frac{M_{\rm gas}}{2\Msol}\right)\left(\frac{\rho_{\rm ISM}}{10^{-18} \gram/\cm^{3}}\right)^{-1}\left(\frac{v^{\rm r}}{10^{4}{\rm km/s}}\right),
\end{align}
where $v^{\rm r}$ is the expansion speed and $\rho_{\rm ISM}$ the density of the background medium. At the same time, this also means that the scaling relations for the homologous expansion found from our simulations would  be applied to the evolution of the homologously expanding part of the collision product, independent of the existence of the background medium.

\subsection{Interaction of gas cloud with supermassive black hole}\label{subsec:bhgas}
In addition to the burst caused by the stellar collision (see \S~\ref{subsec:observables}), there would be a subsequent burst due to accretion onto the nearby SMBH. As a result, the overall shape of the luminosity would be that the stellar collision creates the first peak with $L\gtrsim 10^{42}$ erg/s which decays, followed by a sharp rise to Eddington due to accretion onto the BH, possibly remaining at that level for up to years until the captured gas is accreted onto the BH. We will examine the observables from the BH-cloud interaction by considering two cases, 1) Case 1. no-decelerating expansion (\S\ref{subsubsec:dec}) and 2) Case 2. decelerating expansion (\S\ref{subsubsec:nodec}). For simplicity, we assume that the center of mass motion of the collision product is moving sufficiently slowly compared to the cloud expansion speed, which would be relevant for head-on equal-mass collisions. In addition, we make a crude assumption that the gas cloud is expanding spherically. However, in reality, the collisions product can have a non-negligible coherent motion compared to the SMBH and the shape of the gas cloud would be deformed by the tidal force of the SMBH. These will be studied in detail in a follow-up project. And we will discuss the astrophysical implications for BHs in \S\ref{subsubsec:bhimplication}.

\subsubsection{Case 1. no-decelerating expansion}\label{subsubsec:dec}

We first assume that the entire gas cloud expands homologously and the expansion speed of the outer edge is 
$v^{\rm r}\simeq \psi v_{\rm rel}$ with $\psi\simeq 3-6$ (see Figure~\ref{fig:density_vr2}). The gas cloud starts to interact with the BH when the size of the expanding gas cloud becomes comparable to the distance to the BH $R_{\rm BB}$ for given $v_{\rm rel}$,
\begin{align}
    R_{\rm BH}\simeq \frac{G\Mbh}{v_{\rm rel}^{2}} = 10^{15}~ {\rm cm} \left(\frac{\Mbh}{10^{7}\Msol}\right)\left(\frac{v_{\rm rel}}{10^{4}{\rm km/s}}\right)^{-2}.
\end{align}
The time difference between the first collision-driven burst and the subsequent accretion-driven burst would be set by the time $\tau_{\rm BH}$ at which the outer edge of the cloud reaches the BH, $R_{\rm BH} - R_{\rm Sch}\simeq R_{\rm BH}\simeq R_{\rm peak}$, where $R_{\rm Sch}$ is the Schwarzschild radius,
\begin{align}
    \tau_{\rm BH}\simeq \frac{R_{\rm BH}}{v^{\rm r}}\simeq 3 ~{\rm days} \left(\frac{\Mbh}{10^{7}\Msol}\right)\left(\frac{v_{\rm rel}}{10^{4}{\rm km/s}}\right)^{-3}.
\end{align}
To zeroth order, the part of the cloud that is within the Bondi radius $R_{\rm Bondi}\simeq 2G\Mbh/(v^{\rm r})^{2}$ from the BH would be gravitationally captured by the BH and subsequently accreted onto the BH. Assuming a Bondi–Hoyle accretion \citep{BondiHoyle1944,Bondi1952}, the luminosity $L_{\rm Bondi}$ with radiative efficiency $\epsilon$ can be estimated,
\begin{align}
   &L_{\rm Bondi}\simeq \frac{4\pi \epsilon G^{2}\Mbh^{2}\rho c^{2}}{(v^{\rm r})^{3}},\nonumber\\
    &\simeq3\times 10^{47}~{\rm erg /s} \left(\frac{\epsilon}{0.1}\right)\left(\frac{\Mbh}{10^{7}\Msol}\right)^{-1}\left(\frac{v_{\rm rel}}{10^{4}{\rm km/s}}\right)^{3},
\end{align}
which is super-Eddington for $\Mbh< 3\times 10^{8}\Msol (v_{\rm rel}/10^{4}{\rm km~s}^{-1})^{1.5}$. Note that $L_{\rm Bondi}$ has no dependence on $t$ given the scaling relations for $\rho$ ($\propto t^{-3}$, Equation~\ref{eq:rho}) and $v^{\rm r}(r=R_{\rm BH})$ ($\propto t^{-1}$, Equation~\ref{eq:vr}): $L_{\rm Bondi}\propto \rho (v^{\rm r})^{-3}\propto t^{0}$. Super-Eddington accretion may be possible if the gas is optically thick and the trapping radius $R_{\rm tr}=(L_{\rm Bondi}/L_{\rm Edd})(G M_{\bullet}/\epsilon c^{2})$ is smaller than the Bondi radius \citep{Begelman1979}. The ratio of the two radii is,
\begin{align}\label{eq:tratio}
\frac{R_{\rm tr}}{R_{\rm Bondi}}\simeq 300\left(\frac{t}{1 {\rm day}}\right)^{-2}\left(\frac{v_{\rm rel}}{10^{4}{\rm km/s}}\right)^{-1},
\end{align}
suggesting super-Eddington accretion would be possible at $t\gtrsim 20~{\rm days}(v_{\rm rel}/10^{4}{\rm km s}^{-1})^{-0.5}$.  Here, we caution that the time ratio in Equation~\ref{eq:tratio} is estimated under the assumption that the global accretion flow is not affected by any accretion feedback, which is highly uncertain. Assuming a black body, the temperature at the Bondi radius if $L\simeq L_{\rm Edd}$ is,
\begin{align}\label{eq:Tbondi1}
T_{\rm Bondi}(t)\simeq 10^{5} K \left(\frac{t}{3 {\rm day}}\right)^{-1}\left(\frac{\Mbh}{10^{7}\Msol}\right)^{3/4}\left(\frac{v_{\rm rel}}{10^{4}{\rm km/s}}\right)^{-1/2},
\end{align}
and at the onset of the BH-gas interaction (or $t\simeq \tau_{\rm BH}$),
\begin{align}\label{eq:Tbondi2}
T_{\rm Bondi}(t=\tau_{\rm BH})\simeq 10^{5} K \left(\frac{\psi}{5}\right)\left(\frac{\Mbh}{10^{7}\Msol}\right)^{-1/4}\left(\frac{v_{\rm rel}}{10^{4}{\rm km/s}}\right). 
\end{align}
If $L\simeq L_{\rm Bondi}$,
\begin{align}\label{eq:Tbondi3}
T_{\rm Bondi}&\simeq 2.4\times10^{4} K \left(\frac{t}{150 {\rm day}}\right)^{-1}\left(\frac{\epsilon}{0.1}\right)^{1/4}\nonumber\\
&\times\left(\frac{\Mbh}{5\times 10^{8}\Msol}\right)^{1/4}\left(\frac{v_{\rm rel}}{10^{4}{\rm km/s}}\right)^{-5/4},
\end{align}
and at $t=\tau_{\rm BH}$,
\begin{align}\label{eq:Tbondi4}
T_{\rm Bondi}(t=\tau_{\rm BH})&\simeq 2.4\times10^{4} K \left(\frac{\psi}{5}\right)\left(\frac{\epsilon}{0.1}\right)^{1/4}\nonumber\\
&\times\left(\frac{\Mbh}{5\times 10^{8}\Msol}\right)^{-3/4}\left(\frac{v_{\rm rel}}{10^{4}{\rm km/s}}\right)^{-7/4}.
\end{align}
Because $R_{\rm Bondi}$ increases faster than $R_{\rm peak}$,
\begin{align}
    \frac{R_{\rm Bondi}}{R_{\rm peak}}\propto t,
\end{align}
as the most optimistic case, the entire gas cloud could be ultimately captured by the BH in a time $\tau_{\rm capture}$ at which $R_{\rm Bondi}\simeq R_{\rm peak}$,
\begin{align}
    \tau_{\rm capture}\simeq 40~{\rm days} \left(\frac{\psi}{5}\right)\left(\frac{\Mbh}{10^{7}\Msol}\right)\left(\frac{v_{\rm rel}}{10^{4}{\rm km/s}}\right)^{-3}.
\end{align}
Then the maximum duration of the Eddington luminosity may be set by,
\begin{align}\label{eq:tacc}
    \tau_{\rm acc} \lesssim \frac{M_{\rm gas}\epsilon c^{2} }{L_{\rm Edd}}\simeq 9~{\rm years}\left(\frac{\epsilon}{0.1}\right)\left(\frac{M_{\rm gas}}{2\Msol}\right)\left(\frac{\Mbh}{10^{7}\Msol}\right)^{-1},
\end{align}
where $M_{\rm gas}$ is the mass of the gas cloud, i.e., total mass of the two collided stars. Here, we assumed that the entire gas would be accreted onto the BH. However, radiation pressure from super-Eddington accretion would be strong enough to generate outflow. For such a case, only a fraction of the gas cloud would end up accreting and $\tau_{\rm acc}$ would be shorter than estimated above.

\begin{figure}
	\centering
\includegraphics[width=8.7cm]{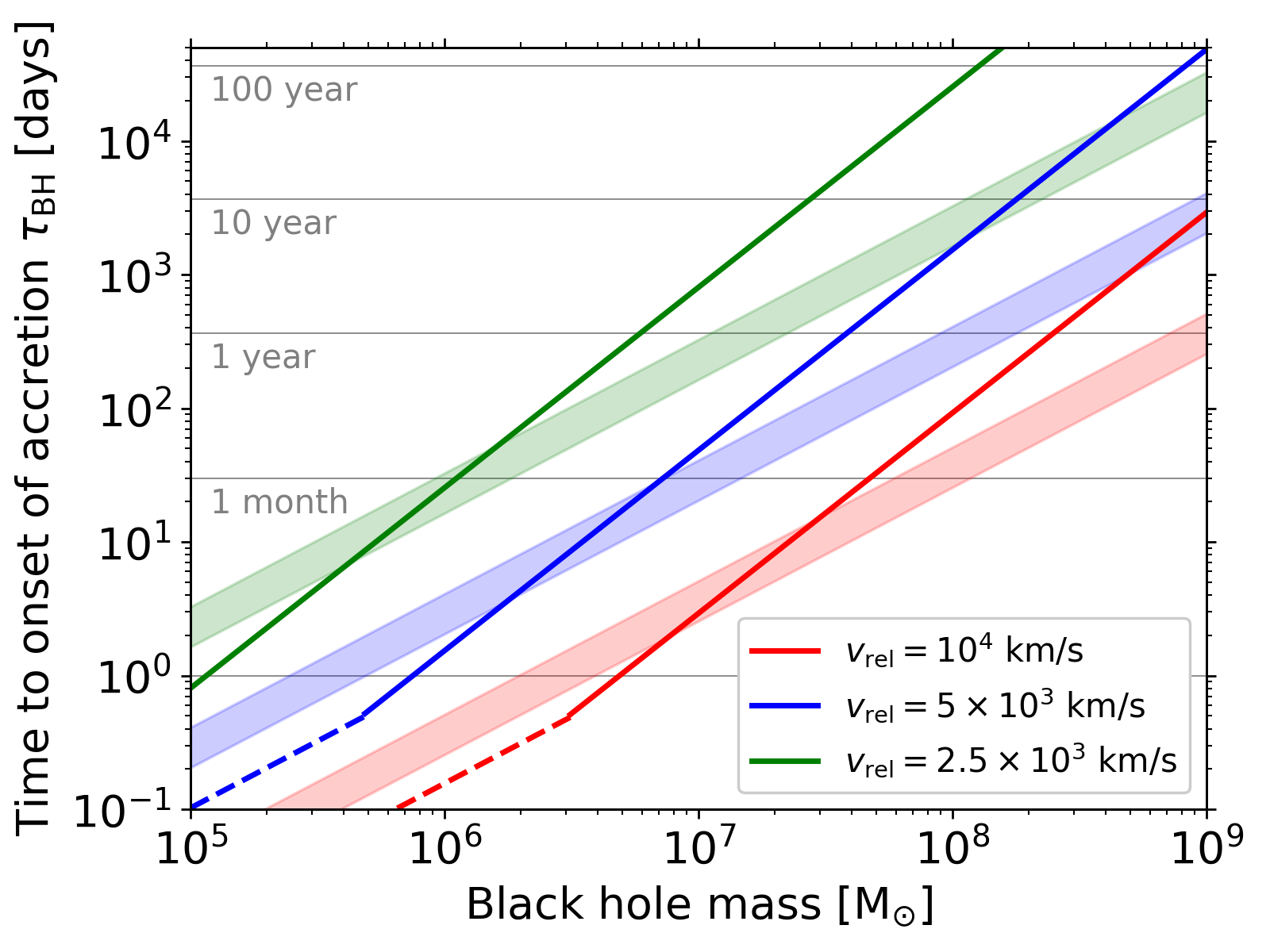}
\caption{Time $\tau_{\rm BH}$ to the accretion-driven burst since the peak collision-driven luminosity for different collision velocities $v_{\rm rel}$, as a function of black hole mass.  The lines illustrate \textit{Case 2. decelerating expansion} (\S~\ref{subsubsec:dec}) where the entire cloud expands homologously up to 0.5 days since collision (dashed), then the outer edge starts decays like $t^{-1/3}$ (solid) due to interactions with a background medium, using Equation~\ref{eq:vr}. The less steep diagonal bars demarcate the range of $\tau_{\rm BH}$ for the case where the gas cloud continuously expands homologously with the outer edge moving at $(3-6)\times v_{\rm rel}$ (\textit{Case 1. no-decelerating expansion} (\S~\ref{subsubsec:nodec}), corresponding to the peak expansion speed upon collision in our simulations (see the \textit{bottom-right} panel of Figure~\ref{fig:density_vr2}).  }
	\label{fig:tauBH}
\end{figure}

\begin{figure}
	\centering
\includegraphics[width=8.6cm]{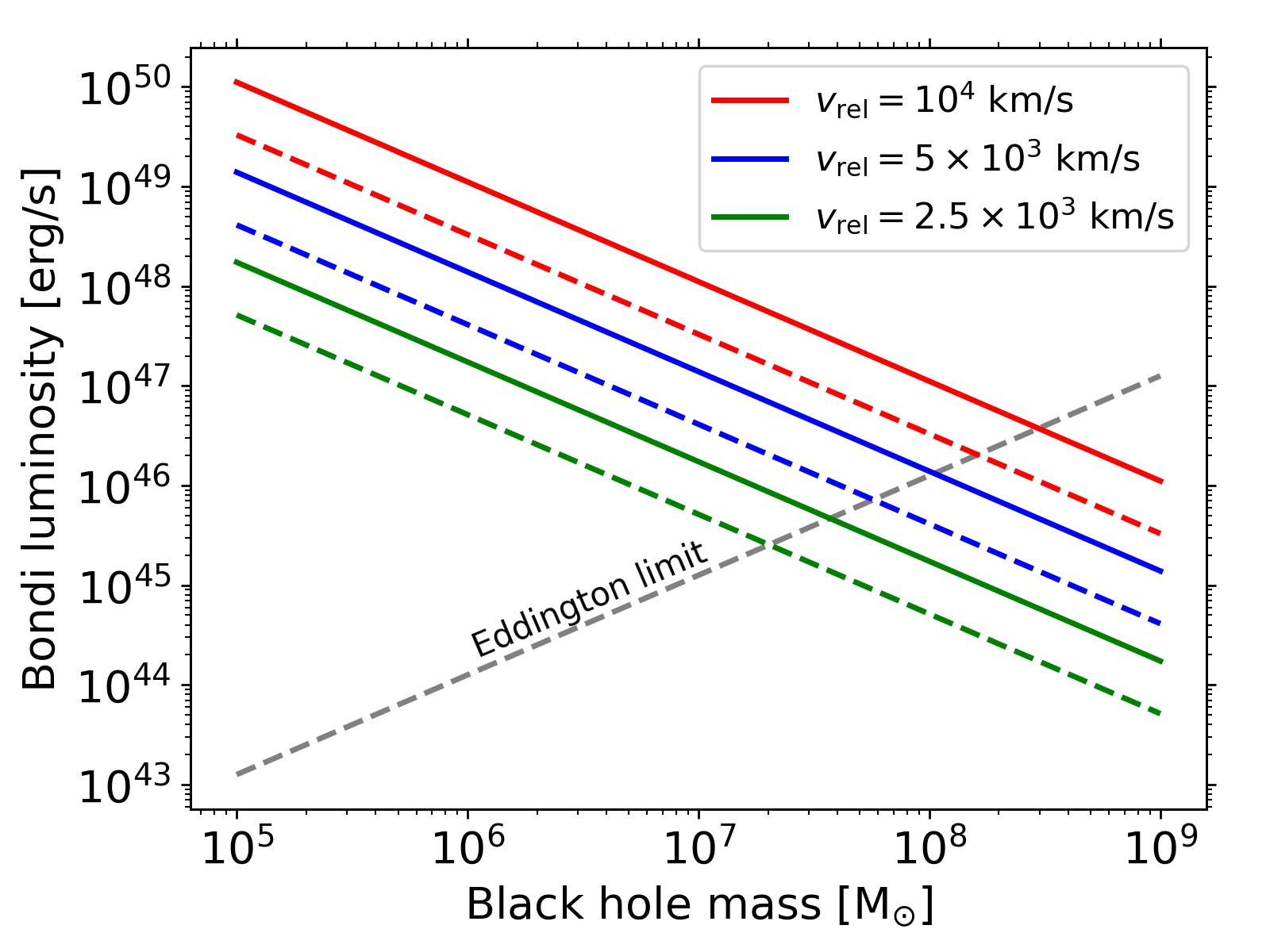}
\caption{Bondi luminosity due to free-fall accretion of the collision product onto a black hole for different collision velocities $v_{\rm rel}$, as a function of black hole mass. The solid line is for the decelerating peak expansion speed (\textit{Case 2. decelerating expansion}, \S~\ref{subsubsec:dec}) and dashed lines for the non-decelerating expansion speed (\textit{Case 1. no-decelerating expansion} (\S~\ref{subsubsec:nodec}). The grey dashed diagonal line indicates the Eddington luminosity.  }
	\label{fig:LBondi}
\end{figure}

\subsubsection{Case 2. decelerating expansion}\label{subsubsec:nodec}
Now we examine the observables from interactions between decelerating expanding cloud with $v^{\rm r}_{\rm peak}\propto t^{-1/3}$ and the SMBH, using Equations~\ref{eq:rho}--\ref{eq:vr}. For this case, $\tau_{\rm BH}$ has a different dependence on $\Mbh$ and $v_{\rm rel}$,
\begin{align}
    \tau_{\rm BH}\simeq 3 ~{\rm days} \left(\frac{\Mbh}{10^{7}\Msol}\right)^{3/2}\left(\frac{v_{\rm rel}}{10^{4}{\rm km/s}}\right)^{-3}\left(\frac{b/R_{\star} + 5}{5}\right)^{6},
\end{align}
We show in Figure~\ref{fig:tauBH} the range of $\tau_{\rm BH}$ for three different collision velocities $v_{\rm rel}$ as a function of $\Mbh$ assuming a non-decelerating expansion speed (thick diagonal bars, $\psi=3-6$) and a decelerating expansion speed (solid lines). The interaction onset time would be longer generally if the expansion of the cloud slows down. Depending on $M_{\rm BH}$ and $v_{\rm rel}$, the second burst could happen over a wide range of time. For example, if a collision with $v_{\rm rel}\gtrsim 2500$ km/s occurs in the Galactic center \citep[with $\Mbh\simeq 4\times10^{6}\Msol$][]{GRAVITY}, the second accretion-driven burst would occur after the collision in less than a day to $6-7$ months depending on the location of the collision from the BH. For very massive black holes ($M_{\rm BH}>10^{8}\Msol$), $\tau_{\rm BH}$ can be more than tens of years. 

\begin{figure}
	\centering
\includegraphics[width=8.5cm]{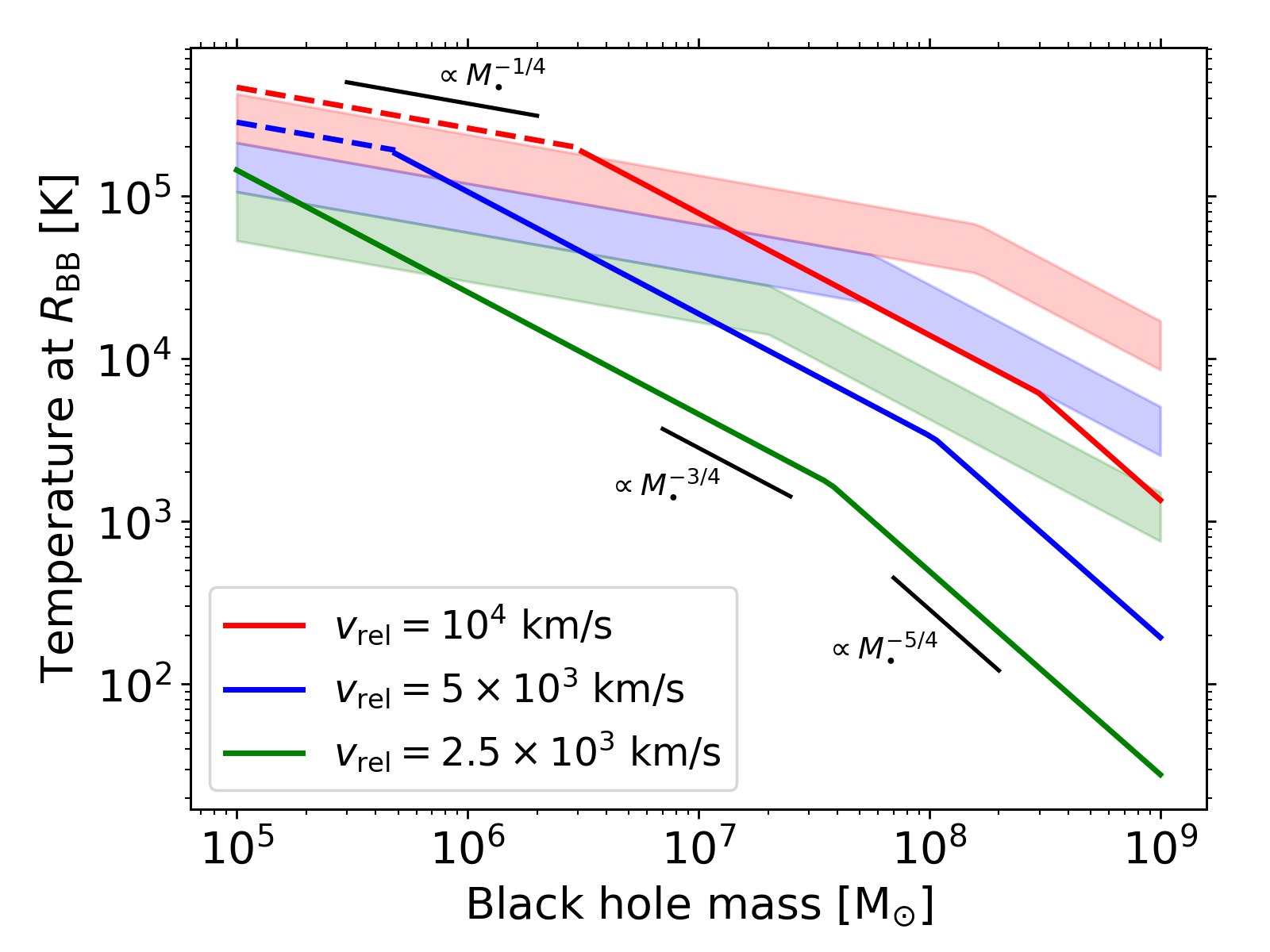}
\caption{Temperature from the Eddington-limited Bondi-luminosity at the onset of the accretion of the collision product onto a black hole for different collision velocities $v_{\rm rel}$, as a function of black hole mass. As before in Figure~\ref{fig:tauBH}, the lines show the case where the cloud undergoes a non-decelering expansion up to 0.5 days, followed by an deceleration of the outer edge like $t^{-1/3}$ at 0.5 days since collision due to interactions with a background medium, using Equations~\ref{eq:rho}-\ref{eq:vr}. The diagonal bars indicate the range of $\tau_{\rm BH}$ when the entire gas cloud expands without being decelerated with the peak expansion speed of $(3-6)\times v_{\rm rel}$. The power-laws are analytically derived in Equations~\ref{eq:Tbondi1} ($\propto \Mbh^{-1/4}$) and \ref{eq:Tbondi5} ($\propto \Mbh^{-3/4}$ and $\propto \Mbh^{-5/4}$). }
	\label{fig:TBondi}
\end{figure}

The Bondi luminosity is still independent of $t$ and has the same $\Mbh-$ and $v_{\rm rel}-$dependence as the case with the no-decelerating expansion, but it is roughly a factor of 3 greater at given $M_{\rm BH}$ and $v_{\rm rel}$,
\begin{align}
   &L_{\rm Bondi}\simeq10^{48}~{\rm erg /s} \left(\frac{\epsilon}{0.1}\right)\left(\frac{\Mbh}{10^{7}\Msol}\right)^{-1}\left(\frac{v_{\rm rel}}{10^{4}{\rm km/s}}\right)^{3},
\end{align}
which is further illustrated in Figure~\ref{fig:LBondi}. While the expression for the blackbody temperature at the Bondi radius has the same dependence on $\Mbh$ and $v_{\rm vel}$ as Equations~\ref{eq:Tbondi1} and ~\ref{eq:Tbondi3}, because of the different expression for $\tau_{\rm BH}$, $T_{\rm Bondi}(t=\tau_{\rm BH})$ is written differently,
\begin{align}\label{eq:Tbondi5}
&T_{\rm Bondi}(t=\tau_{\rm BH})\simeq\nonumber\\
&\begin{cases}
   8\times10^{4}K \left(\frac{\Mbh}{10^{7}\Msol}\right)^{-3/4}\left(\frac{v_{\rm rel}}{10^{4}{\rm km/s}}\right)^{2.1}\left(\frac{b/R_{\star} + 5}{5}\right)^{-6},\\
    \hspace{2.5in}\textrm{for $L=L_{\rm Edd}$}, \\ 
  6\times10^{3}K \left(\frac{\epsilon}{0.1}\right)^{1/4} \left(\frac{\Mbh}{10^{7}\Msol}\right)^{-5/4}\left(\frac{v_{\rm rel}}{10^{4}{\rm km/s}}\right)^{2.8}\left(\frac{b/R_{\star} + 5}{5}\right)^{-6},\\
   \hspace{2.5in}\textrm{for $L=L_{\rm Bondi}$}. \\ 
\end{cases}
\end{align}
We compare $T_{\rm Bondi}$ at the onset of the accretion-drive burst (so $T_{\rm Bondi}$ at $t=\tau_{\rm BH}$) in Figure~\ref{fig:TBondi} between the non-decelerating expansion case (thick bars) and the decelerating expansion case (lines). For low-mass black holes, $T_{\rm Bondi}$ is quite similar, e.g., $10^{5}$ K for $\Mbh=10^{5}-10^{6}\Msol$. However, because of a steeper decline for the decelerating expansion case ($T_{\rm Bondi}\propto \Mbh^{-3/4}-\Mbh^{-5/4}$, Equation~\ref{eq:Tbondi5}) than for the no-decelerating expansion case ($T_{\rm Bondi}\propto \Mbh^{-1/4}-\Mbh^{-3/4}$, Equations~\ref{eq:Tbondi2} and \ref{eq:Tbondi4}), $T_{\rm Bondi}$ for the decelerating expansion case is generally lower for high-mass BHs: for $\Mbh=10^{9}\Msol$, $T_{\rm Bondi}\simeq 10 - 10^{3}$ K for the decelerating expansion case whereas $T_{\rm Bondi}\simeq 10^{3}-10^{4}$ K for the no-decelerating expansion case.

For the decelerating expansion case, the Bondi radius increases faster,
\begin{align}
    \frac{R_{\rm Bondi}}{R_{\rm peak}}\propto t^{4/3},
\end{align}
which leads to a smaller $\tau_{\rm capture}$,
\begin{align}
    \tau_{\rm capture}\simeq 11 ~{\rm days} \left(\frac{\Mbh}{10^{7}\Msol}\right)^{3/4}\left(\frac{v_{\rm rel}}{10^{4}{\rm km/s}}\right)^{-2.5}\left(\frac{b/R_{\star} + 5}{5}\right)^{-3}.
\end{align}
The duration of the accretion process would be the same as Equation~\ref{eq:tacc}.

\begin{figure}
	\centering
\includegraphics[width=8.5cm]{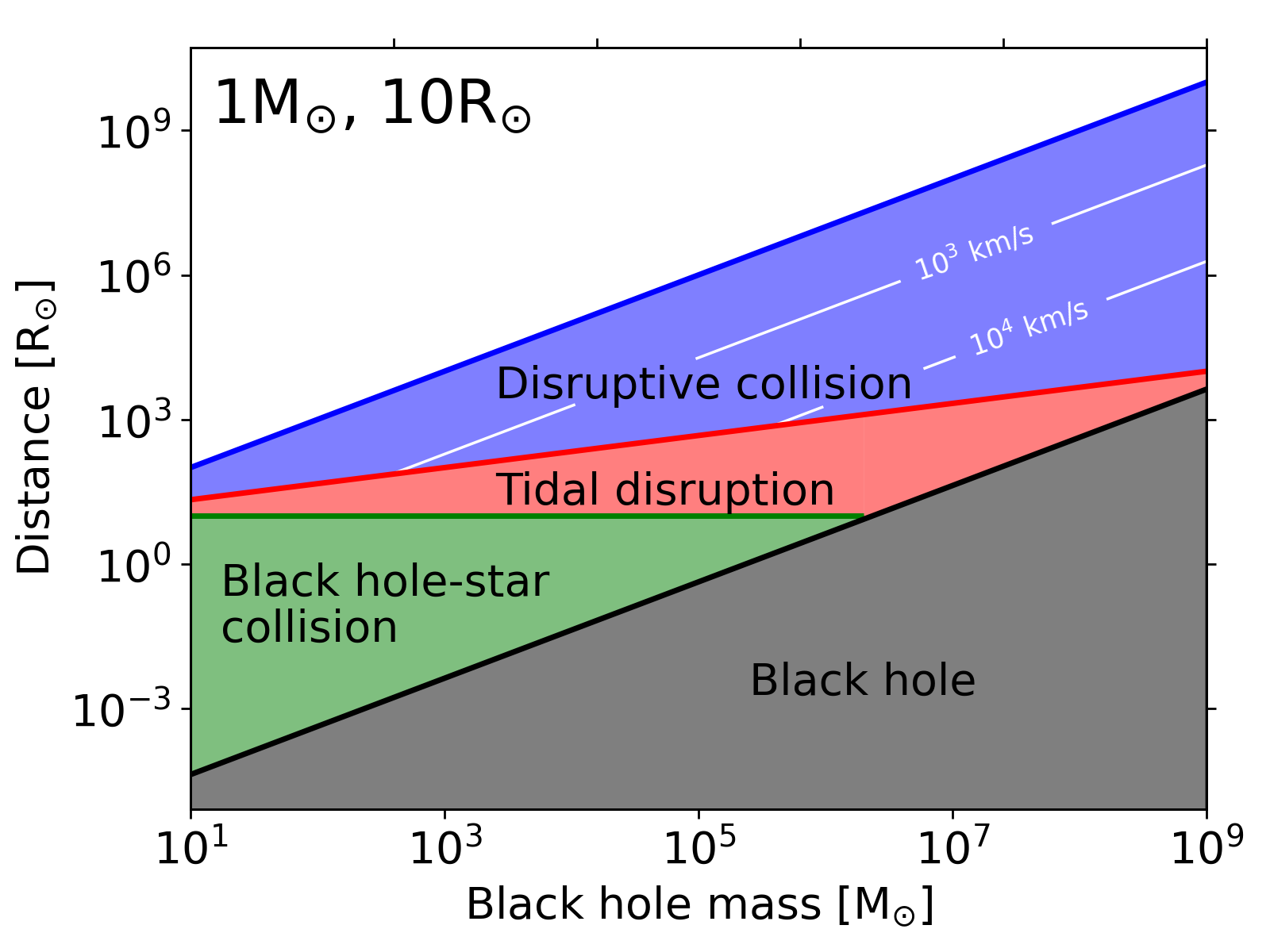}
\caption{Parameter space for disruptive events of a giant with $\mstar=1\Msol$ and $\rstar=10\Rsol$ in terms of the distance from the BH for varying BH masses. The region dubbed ``Black hole`` is defined by the Schwarzschild radius, $r_{\rm Sch}=2 G\Mbh/c^{2}$. If the pericenter distance of a star is smaller than a few times outside $r_{\rm Sch}$, the star would be directly captured by the black hole. If the separation is smaller than the stellar radius, they would collide (``black hole-star collision''). When the pericenter distance is smaller than the tidal radius $r\leq r_{\rm t}= (M_{\bullet}/M_{\star})^{1/3}R_{\star}$, stars are tidally destroyed by the BH (``tidal disruption''). Finally, disruptive collisions happen between the distance at which the Keplerian velocity is greater than the stellar escape velocity, $r\leq r_{\rm collision}= (M_{\bullet}/M_{\star})R_{\star}$ and the tidal radius $r\leq r_{\rm t}$. The white diagonal lines in the region for disruptive collisions corresponds to the collision velocity for given BH mass and radius.}
	\label{fig:distance}
\end{figure}

\subsubsection{Astrophysical implication for black holes }\label{subsubsec:bhimplication}

The possibility of the accretion of at least some fraction of the expanding cloud onto the SMBH in proximity can have significant implications for the growth of BHs in the cosmic landscape. While several mechanisms for massive BH formation have been proposed, the precise mechanism for growing BH seeds at extremely high redshifts remains uncertain \citep[see for reviews][]{ColpiDotti2011ASL.....4..181C,Inayoshi+2020}. The proposed mechanisms include 1) rapid growth of the remnants of the Population III stars via super-Eddington accretion \citep[e.g.,][]{Volonteri+2005,HaimanLeob2001,Ryu+2016,Lupi+2016,Sassano+2023}, the direct collapse of supermassive self-gravitating objects \citep[e.g.,][]{OmukaiNishi1998,Yoshida+2008,2023MNRAS.518.2076Z}, growth of BHs in a runaway process \citep[e.g.,][]{Devecchi+2012,Stone+2017,Tagawa+2020,Rizzuto+2023}. In principle, as long as a BH is more massive than colliding stars, the velocity of stars around the BH can be large enough that stellar collisions can be disruptive. Hence, the accretion of gas produced in stellar collisions onto a nearby BH can provide another venue for the growth of stellar-mass BHs to massive BHs, in particular see BHs at high redshift.  

However, disruptive collisions are not the only growth mechanism for BHs in stellar-dense environments. We show in Figure~\ref{fig:distance} the regions around BHs in which several events possibly contributing to their growth, i.e., disruptive collisions, tidal disruption events, BH-star collisions, and direct captures by BHs, can occur. When the distance from the BH is less than a few times greater than the Schwarzschild radius $r_{\rm Sch} = 2 G\Mbh/c^{2}$ (dubbed ``direct capture'' radius), the star would directly fall into the BH (e.g., $r<2r_{\rm Sch}$ for parabolic orbits). If the closest approach distance between the BH and a star is smaller than the stellar radius, $r\lesssim\rstar$, they collide, during which the BH would gravitationally capture a fraction of the star and accrete. When a star orbits at a distance greater than both the stellar radius and the direct capture radius, and smaller than the so-called tidal radius, $r_{\rm t} = (M_{\bullet}/M_{\star})^{1/3}R_{\star}$, very strong BH's tidal forces disrupt the star, creating debris, some of which would end up accreting onto the BH. This event is called tidal disruption event \citep{Hills1988,Rees1988}. Finally, the region for disruptive collisions between giants may be characterized by the two distances, the distance within which the Keplerian velocity around the BH exceeds the stellar escape speed, $r_{\rm collision}=(M_{\bullet}/M_{\star})R_{\star}$ and the tidal radius. 

As shown in Figure~\ref{fig:distance}, all four star-destroying events can contribute to the growth of stellar-mass and intermediate-mass BHs. However, for SMBHs with $r_{\rm Sch}> \rstar$, only three events, namely, disruptive collisions, tidal disruptions, and direct captures, can feed the BHs. For very massive BHs (e.g., $M_{\bullet}>10^{9}\Msol$), disruptive collisions would be the dominant and likely only observable transient among those considered here that lead to the mass growth of the BHs \citep[see][]{AmaroSeoane2023b}.

This has an interesting implication for the detection of dormant BHs. TDEs have been considered a unique signpost for the existence of dormant SMBHs. However, because there is a maximum BH mass capable of disrupting stars, $M_{\bullet}$ at which $r_{\rm t}$ equals to the direct capture radius, TDEs can not be used to detect very massive quiescent black holes. However, disruptive stellar collisions can occur near BHs at all mass scales, which would make these events \textit{a promising tool to probe the existence of very massive dormant BHs }which cannot be probed by other transients. In particular, if the luminosity due to the interaction of the collision product with the BH is Eddington-limited, an inference of the BH mass would potentially be possible from the observed radiated light curve.

Which type of the events is dominant at different mass ranges would depend on the stellar density, the accretion efficiency and occurrence rates, which is beyond the scope of our paper. We will examine this aspect in more detail in our future work.

\subsection{Particle acceleration}

In this work we have conducted numerical hydrodynamical simulations that confirm, following stellar collision event, the formation of strong shocks. These shocks arise due to the high velocity of the outflow, and its impact with the surrounding ISM in the galactic nucleus environment. These shock waves subsequently compress and heat the surrounding ISM gas.

The shocks formed in these stellar collisions provide an environment highly conducive to efficient particle acceleration. As particles interact with the turbulent magnetic fields expected close to the shock front, they can gain a significant fraction of the free energy available from the differential flow speeds (in the shock's rest frame, the upstream towards the shock with velocity $V$ and the downstream moves away from the shock at velocity $V/4$). This process of diffusive particle acceleration at shocks, an example of first order Fermi acceleration, is expected to result in the generation of a power-law spectrum of non-thermal particles up to very high energies \citep[]{1978ApJ...221L..29B,1978MNRAS.182..147B}. 

A fraction of the energy in the accelerated particle population produced by stellar collisions will subsequently be radiated via non-thermal emission through various energy loss processes \citep[see ][for revieww in the context of active galactic nuclei jets and supernovae, respectively]{Matthews+2020,Orlando+2021}. 
For instance, the accelerated electrons will produce synchrotron radiation as they spiral around the magnetic fields also generated during the collision. This emission is expected to be detectable in the radio, and potentially the X-ray, bands. In addition, the interaction between accelerated protons and the surrounding gas can generate gamma-ray emission through processes like inelastic proton-proton collisions.

The non-thermal radiation emitted by the accelerated particles produced in stellar collisions offers valuable diagnostics into the physical processes at play during violent stellar collision events. By analyzing the observed non-thermal radiation, we can gain a clearer understanding of shock front environment. Ultimately, these insights will elucidate on the dynamics of the collision itself. Our numerical hydrodynamics simulations, coupled with theoretical estimates for the production of non-thermal particles not included in our numerical description, provide can provide insights into particle acceleration in stellar collisions. This will be addressed elsewhere in a separate work.

\subsection{Destructive collisions between different types of stars}
Although we only consider BDCs between equal-mass $1\Msol$ giants, there could be a variety of BDCs involving various types of stars. The total radiated energy, luminosity, and temperature of BDCs would be affected by various factors, including the relative size and mass of the colliding stars.

 It is essential to convert the collision kinetic energy into radiation energy for generating bright flares. For efficient energy conversion, one important requirement is a large contact area at collision. On the one hand, when two stars with two significantly different radii collide (e.g., main-sequence and giant) at high velocity, even if the impact parameter is small, the smaller star would simply penetrate through the fluffy envelope of the larger star. On the other hand, for collisions involving stars with comparable sizes, the dependence of luminosity on the radii of the two colliding stars may not be so simple because how strong shocks are created at collision and how rapidly photons escape after collision would be determined by several factors, such as the temperature and density of the envelop of the star before collision and those of the gas cloud after collision. At least, our simulations suggest that collisions involving larger giants are brighter because of stronger collision shocks over a wider cross section (see Figures~\ref{fig:observables} and \ref{fig:observables2}). In addition, the thermodynamic properties and chemical elements of the gas cloud would affect the spectra of the flare \citep{Dessart+2023}. 

Another important factor that determines the collision outcome and the amount of radiated energy is the masses of the colliding stars. 
For the cases where two stars are completely destroyed due to collision, the total energy radiated away would be limited by the total mass of the stars. However, it is also possible that only one of the two colliding stars is completely destroyed. For such cases, while the total energy radiated away due to the collision would be limited by the mass of the star that is destroyed, the relatively intact star may play a role as an extra energy source that may affect the long-term evolution of the light curves. In addition, unlike the cases considered in this paper, the radiation would be highly asymmetric.

Taking all these factors into account, flares due to BDCs would generally be brighter when two stars with larger masses and comparable sizes collide at a higher velocity with a smaller impact parameter. However, due to the shorter lifetime and lower number density of more massive stars, the collision rates would be generally smaller.   

\section{Conclusion and Summary}\label{sec:summary}

In this work, we investigate the hydrodynamics of high-velocity collisions between giants in galactic nuclei and their observational signatures using two state-of-the-art codes, the 3D moving-mesh hydrodynamics code {\small AREPO} and the 1D stellar evolution code {\small MESA}. The initial conditions of our simulations involved two identical $1\Msol$ giants with different radii, initial relative speeds, and impact parameters. This work complements to the analytical calculations presented by \citet{AmaroSeoane2023}, which is generally consistent with each other. We improve the estimates of the events' observables by accurately taking into account the realistic stellar internal structure and non-linear hydrodynamics effects.

When two stars collide with exceedingly large kinetic energy, very strong shocks are created along the contact surface. The  envelope of the two giants are fully destroyed and merged into an homologously, quasi-spherically, and supersonically expanding gas cloud. The maximum expansion speed of the cloud is larger than the initial relative velocity of the stars by a factor of $3-6$. The expansion speed at a given mass coordinate stays the same, but the outer edge of the cloud slows down because of the interaction with the background medium. As it expands, the overall level of its density and temperature drops following a power-law $\propto t^{-3}$ and $\propto t^{-1}$, respectively, becoming optically thin within a few hundred days. At any given time of evolution up to 30 days, the density and temperature of the inner regions of the cloud remain relatively constant, rapidly decaying towards the outer edge, following a power-law: $\rho(r)\propto r^{-8}-r^{-12}$ and $T(r)\propto r^{-1} - r^{-2}$.  These quantities exhibit weak dependencies on the stellar radius within $10-100\Rsol$ and the impact parameter within $b\simeq 0.4\Rsol$. But the dependence on the collision velocity is relatively strong. We provide fitting formulae for the average cloud density, temperature, maximum expansion speed, and optical depth (Equations~\ref{eq:rho}- \ref{eq:vr}), which would be useful for analytic estimates for these high-velocity stellar collisions.

One of the key findings of our study is to numerically estimate the amount of radiation energy converted from the initial kinetic energy, which plays a crucial role in determining the observable properties of the collisions. The overall trend of the conversion efficiency, defined as the ratio of the converted radiation energy to the initial kinetic energy, is such that it peaks at $\gtrsim 0.1$ at collision, decays to $10^{-4} - 10^{-2}$ within 10 days, and then gradually increases. The efficiency reaches $10^{-2} - 10^{-1}$ in one month since the collision. But its magnitude depends on various factors, including the stellar radius, impact parameter, and collision velocity. More specifically, a collision between larger stars colliding at a higher speed with a smaller impact parameter tends to result in greater conversion efficiency. 

We estimate the luminosity, the blackbody radius, and the blackbody temperature, using the converted radiation energy and local cooling time within the gas cloud. The peak luminosity can reach values exceeding $10^{42}$ erg/s and exhibits the similar dependence wtih the conversion efficiency. Over time, the luminosity decays following a power-law of $t^{-0.8}$ at early times and $t^{-0.4}$ after 10 days since collision. The blackbody radius increases almost linearly with time ($\propto t^{0.8}$), while the temperature decreases, following a power-law of $t^{-0.5}-t^{-0.6}$. The collision events would initially produce bursts of extreme ultraviolet ($\simeq10$eV) gradually shifting to optical ($\simeq 0.1$eV), with temporal evolution spanning from days to weeks. These events can be observed by ongoing (e.g., ZTF  \citealt{ZTF}\footnote{https://www.ztf.caltech.edu}, ASSA-SN \citealt{ASASSN}\footnote{https://www.astronomy.ohio-state.edu/asassn}) and future (e.g., LSST \citealt{LSST}\footnote{https://www.lsst.org} and ULTRASAT \citealt{ULTRASAT}\footnote{https://www.weizmann.ac.il/ultrasat}) surveys. More detailed radiation transport calculations will be carried out in our follow-up project, with which the detection rate for each survey will be estimated.

In addition to the burst resulting from the stellar collision itself, a subsequent burst occurs due to the accretion of the gas cloud onto the supermassive black hole in the galactic center in $5(M_{\bullet}/10^{7}\Msol)$ days for $v_{\rm rel}=10^{4}$ km/s since collision. Assuming Bondi accretion, the accretion luminosity can easily exceed the Eddington limit as well as the luminosity from the stellar collision. Because the Bondi radius expands faster than the gas cloud, the entire cloud would be gravitationally captured in the black holes's potential in $11(M_{\bullet}/10^{7}\Msol)^{3/4}$ days and subsequently accrete onto the black hole. It would take $\lesssim 9 (M_{\bullet}/10^{7}\Msol)^{-1}$ years if the entire cloud was accreted. Therefore, the overall luminosity curve would include a peak from the collision event, followed by a rise to the Eddington luminosity. This heightened luminosity can be sustained for up to 10 years. 

Although the estimate of the time scales and luminosity due to gas-black hole interactions are still on the order-of-magnitude level, this aspect indicates very important implications. The possibility of the gas accretion onto the black hole at all mass scales in proximity subsequently after the collision suggests that the collision can provide another mechanism for black hole growth. Tidal disruption events have been proposed as a tool to detect dormant black holes, mostly up to $10^{8}\Msol$. However, because disruptive stellar collisions can occur near very massive dormant ones ($>10^{9}\Msol$), such collisions can be a potentially promising tool to probe the existence of very massive dormant black holes.

Finally we demonstrate the conversion of kinetic energy into radiation energy, providing insights into the efficiency of particle acceleration in these collisions. The resulting bursts of ultraviolet and optical emission indicate the generation of high-energy particles, highlighting the importance of particle acceleration processes in understanding the observational signatures of such events.

While this study, to our knowledge, is the first detailed hydrodynamics calculations of high-velocity disruptive collisions between giants, there are a few caveats in our modelling that will be improved in our future work. First, the assumption for local thermodynamic equilibrium is only valid for optically thick gas. This means the evolution of the collision product at early times is accurate, but as the gas cloud becomes optically thin, our treatment of radiation pressure becomes inaccurate. As remarked in \S~\ref{subsec:observables}, This would affect the shape of the lightcurves at late times. We will perform detailed non-equilibrium radiation transport calculations for the late time evolution in our follow-up project using our hydrodynamics calculations at early times when our assumption for local thermodynamic equilibrium is valid. This will significantly improve the light curve modelling. Second, there are several physical impacts that we have not considered yet, such as, magnetic fields, recombination, and the existence of non-thermal particles. Using the machinery that we built for this work, We will explore their impacts in a series of studies dedicated to investigating the impact of each physics. 

The high-velocity disruptive collisions will offer insights into many astrophysical aspects that can not be provided by other transients, such as the stellar dynamics and potential particle acceleration in galactic nuclei and globular clusters, black hole growth, and detection of dormant black holes. 

\section*{Acknowledgements}
The authors are grateful to the anonymous referee for constructive comments and suggestions. TR is grateful to Luc Dessart and Re'em Sari for constructive comments on the manuscript, Hans-Thomas Janka for fruitful discussions for similarities and dissimilarities of these events with core-collapse supernovae, and Ruggero Valli for providing an \mesa{} input file for creating the giants used for the simulations.
This research project was conducted using computational resources (and/or scientific computing services) at the Max-Planck Computing \& Data Facility. The authors gratefully acknowledge the scientific support and HPC resources provided by the Erlangen National High Performance Computing Center (NHR@FAU) of the Friedrich-Alexander-Universität Erlangen-Nürnberg (FAU) under the NHR project b166ea10. NHR funding is provided by federal and Bavarian state authorities. NHR@FAU hardware is partially funded by the German Research Foundation (DFG) – 440719683. In addition, some of the simulations were performed on the national supercomputer Hawk at the High Performance Computing Center Stuttgart (HLRS) under the grant number 44232. PAS acknowledges the funds from the ``European Union NextGenerationEU/PRTR'',
Programa de Planes Complementarios I+D+I (ref. ASFAE/2022/014).

\section*{Data Availability}
Any data used in this analysis are available on reasonable request from the first author.

\bibliographystyle{mnras}

\appendix

\section{Luminosity estimate}\label{appen1}

\begin{figure}
	\centering
\includegraphics[width=8.5cm]{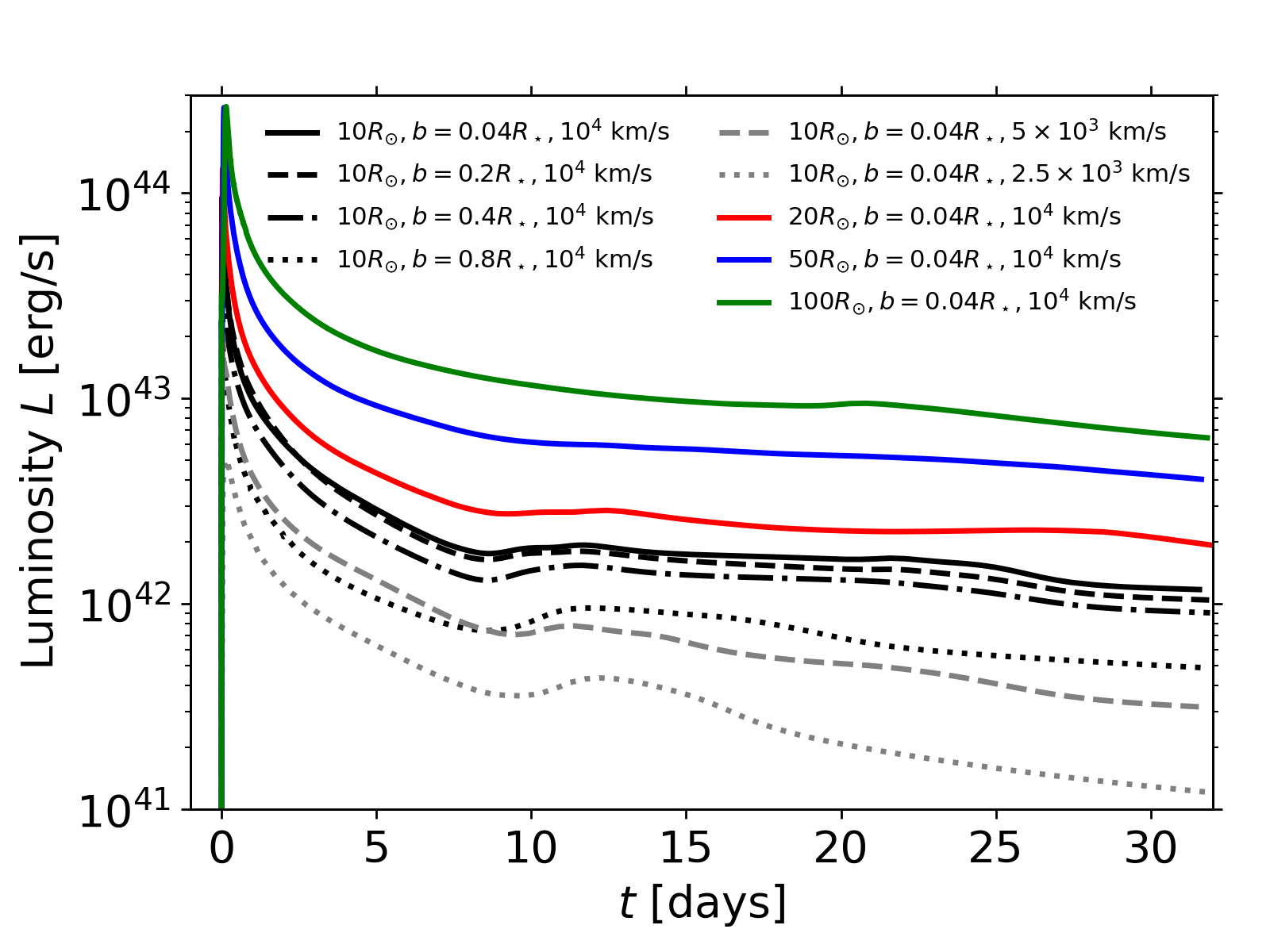}
\includegraphics[width=8.5cm]{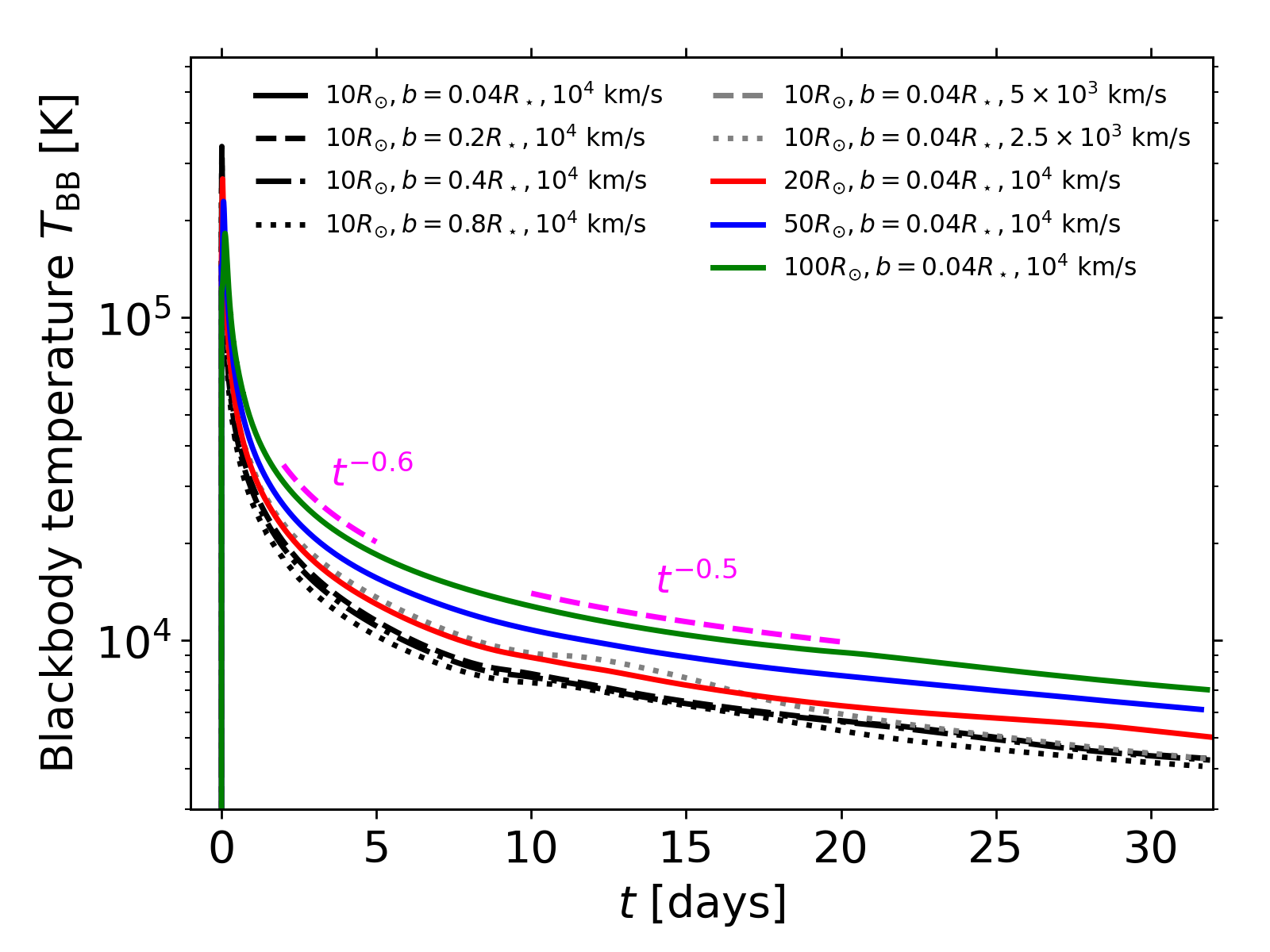}
\caption{Bolometric luminosity $L_{2}$ (\textit{top}) and blackbody temperature $T_{\rm BB}$ (\textit{bottom}) using Equations ~\ref{eq:Abb} and \ref{eq:L2}. As in Figure~\ref{fig:observables}, the magenta guide lines in the \textit{bottom} panel show the power-law that describes the quantity. }
	\label{fig:observables2}
\end{figure}

Figure~\ref{fig:observables2} show the luminosity $L_{2}$ (\textit{top}) estimated using  Equation ~\ref{eq:L2} and the resulting blackbody temperature $T_{\rm BB}$ (\textit{bottom}), as a function of time measured since collision for all our models. 

\end{document}